\documentclass[11pt]{article}

\usepackage{amssymb}
\def\lddots{\mathinner{\mkern1mu\raise1pt\hbox{.}\mkern2mu
\raise4pt\hbox{.}\mkern2mu\raise7pt\vbox{\kern7pt\hbox{.}}\mkern1mu}}
\makeatletter
\def\numberbysection{\@addtoreset{equation}{section}
\def\theequation{\thesection.\arabic{equation}}}
\makeatother

\numberbysection

\newcommand{\be}{\begin{eqnarray}}
\newcommand{\ee}{\end{eqnarray}}
\newcommand{\non}{\nonumber}

\newcommand{\finproof}{{\hfill \rule{5pt}{5pt}}}

\begin{document}

\begin{titlepage}

\vskip 0.4cm

\strut\hfill

\vskip 0.8cm

\begin{center}


{\LARGE From affine Hecke algebras to boundary symmetries}

\vspace{10mm}

{\Large Anastasia Doikou\footnote{e-mail: doikou@lapp.in2p3.fr}}

\vspace{14mm}

\emph{ Laboratoire d'Annecy-le-Vieux de Physique Th{\'e}orique\\
LAPTH, CNRS, UMR 5108, Universit{\'e} de Savoie\\
B.P. 110, F-74941 Annecy-le-Vieux Cedex, France}

\end{center}

\vfill

\begin{abstract}

Motivated by earlier works we employ appropriate realizations of the affine Hecke algebra and we recover previously known non-diagonal solutions of the reflection equation for the $U_{q}(\widehat{gl_n})$ case. The corresponding $N$ site spin chain with open boundary conditions is then constructed and boundary non-local charges associated to the non-diagonal solutions of the reflection equation are derived, as coproduct realizations of the reflection algebra. With the help of linear intertwining relations involving the aforementioned solutions of the reflection equation, the symmetry of the open spin chain with the corresponding boundary conditions is exhibited, being essentially a remnant of the $U_{q}(\widehat{gl_{n}})$ algebra. More specifically, we show that representations of certain boundary non-local charges commute with the generators of the affine Hecke algebra and with the local Hamiltonian of the open spin chain for a particular choice of boundary conditions. Furthermore, we are able to show that the transfer matrix of the open spin chain commutes with a certain number of boundary non-local charges, depending on the choice of boundary conditions.

\end{abstract}

\vfill

\rightline{LAPTH-1066/04}
\rightline{September 2004}

\baselineskip=16pt
\end{titlepage}

\section{Introduction}

One of the great achievements in the domain of quantum integrability is the formulation of the quantum group approach, a method initiated in \cite{kure}--\cite{jimbo2} and used for solving a set of algebraic constraints known as the Yang--Baxter equation. Yang--Baxter equation provides the main framework for formulating and solving integrable field theories on the full line \cite{zamo} and discrete integrable models with periodic (or twisted) boundary conditions \cite{FT, korepin}.
Let $R(\lambda)$ acting on ${\mathbb V}^{ \otimes 2}$, satisfy the Yang-Baxter equation (on ${\mathbb V}^{ \otimes 3}$) \cite{korepin, baxter} \be
R_{12}(\lambda_{1}-\lambda_{2})\ R_{13}(\lambda_{1})\
R_{23}(\lambda_{2}) =R_{23}(\lambda_{2})\ R_{13}(\lambda_{1})\ R_{12}(\lambda_{1}-\lambda_{2}), \label{YBE} \ee then according to \cite{kure}--\cite{jimbo2} the $R$ matrix may be determined by solving a system of linear intertwining relations involving the generators of quantum algebras \cite{jimbo, jimbo2, drinf, tak, cha} and $R$. The fact that these intertwining relations are linear simplifies drastically the computation of solutions of (\ref{YBE}), making the quantum group approach a powerful scheme for solving the Yang--Baxter equation. However, it was also realized in \cite{jimbo} that the $R$ matrix may be written in terms of generators of the  Hecke algebra \cite{hecke, hecke1}. An interesting observation within this context is that one may consider a different route, that is acquire solutions of the Yang--Baxter equation by employing realizations of the Hecke algebra. This is actually the logic that we try to convey in this work. The reasons why we advocate the Hecke algebraic approach are 1) because it provides a universal approach for solving the Yang--Baxter equation as will become clear in the following section, and 2) more importantly because this method bears a rather richer variety of solutions compared to the quantum group approach (see e.g. \cite{mawo, doma}). It should be emphasized however that the existence of the intertwining relations is of great significance in any case, mainly because it allows the systematic study of the underlying symmetry of the corresponding integrable system \cite{kor}--\cite{doikou2}.

An analogous scenario applies in the case of integrable systems with general boundaries. More specifically, in addition to the Yang--Baxter equation one more key equation should be implemented, namely the reflection equation. Let $K(\lambda)$ acting on ${\mathbb V}$ be a solution of the reflection equation (on ${\mathbb V}^{\otimes 2}$) \cite{cherednik, sklyanin},
\begin{equation}
R_{12}(\lambda_{1}-\lambda_{2})\ K_{1}(\lambda_{1})
R_{21}(\lambda_{1}+\lambda_{2})\
K_{2}(\lambda_{2})= K_{2}(\lambda_{2})\ R_{12}(\lambda_{1}+\lambda_{2})\ K_{1}(\lambda_{1})\ R_{21}(\lambda_{1}-\lambda_{2}), \label{re} \end{equation} where $R$ satisfies the Yang--Baxter equation (\ref{YBE}).
It was recently realized that a generalized  quantum group approach can be somehow implemented  for solving the reflection equation. In particular, boundary non--local charges were constructed for the first time in \cite{mene} in the context of the boundary sine-Gordon model in the `free fermion' point. Similarly, in \cite{dema} boundary non-local charges were derived for the affine Toda field theories with certain boundaries conditions, and the corresponding $K$ matrices were attained. The $K$ matrix can be also deduced by solving linear intertwining relations between the $K$ matrix and the reflection algebra generators, defined by (\ref{re})  (see e.g. \cite{dema}--\cite{dege}). Note that it as was recently realized in \cite{pascal2} that the boundary non-local charges associated to $U_{q}(\widehat{sl_{2}})$ case generate the so called tridiagonal algebra.

As proposed in \cite{doma, male} an effective way of finding solutions of the reflection equation (\ref{re}) is by exploiting the existence of certain quotients of the affine Hecke algebra \cite{ahecke} such as e.g. the blob algebra \cite{mawo, doma, male, masa, tl}, formulating the affine Hecke algebraic method \cite{mawo,doma, male, masa}. The boundary analogue of the quantum group approach \cite{mene}--\cite{pascal2} is not rigorously established yet for the $U_{q}(\widehat{gl_{n}})$ case with boundary conditions arising from (\ref{re}). It is therefore desirable to attain solutions of the reflection equation via other effective means, such as the affine Hecke approach \cite{mawo,doma, male, masa}, and then derive systematically generators of the reflection algebra. In other words the Hecke algebraic method allows a rigorous construction of realizations of generators of the reflection algebra, a fact that makes this process quite appealing, further justifying our initial choice of this scheme. 

In this investigation we focus on a special class of integrable lattice models, the integrable quantum spin chains. In the first part we review how solutions of the Yang--Baxter equation arise using certain representations of the Hecke algebra. Then the algebraic monodromy matrix and the corresponding closed $U_{q}(\widehat{gl_n})$ spin chain\footnote{By $U_{q}(\widehat{gl_n})$ spin chain we mean the model built with the $R$ matrix associated to the $U_{q}(\widehat{gl_{n}})$ algebra.} with $N$ sites are constructed. The asymptotic behaviour of the monodromy matrix is examined yielding  coproducts of $U_{q}(\widehat{gl_n})$. In the second part, motivated by earlier works \cite{doma, male, masa}, we present the affine Hecke algebraic method for solving the reflection equation for the $U_{q}(\widehat{gl_n})$ case. Then by deriving a non-diagonal representation of the affine Hecke algebra we recover known solutions of the reflection equation for the $U_{q}(\widehat{gl_n})$ case, originally found in \cite{abad}. It is clear that different representations of the affine Hecke algebra lead to different solutions of the (\ref{re}), however here such an exhaustive study of the various representations and the corresponding solutions is not pursued. Once having specified the $c$--number solution we construct a generalized solution of the reflection equation denoted by ${\cal T}$, from which the transfer matrix of the open spin chain is entailed. The asymptotic behaviour of ${\cal T}$ is examined bearing boundary non-local charges, which are coproduct realizations of the reflection algebra (see also \cite{doikou}, \cite{mene}--\cite{pascal2}). Furthermore, we exploit the existence of linear intertwining relations between the solutions of the reflection equation and the elements of the reflection algebra in order to exhibit the symmetry of the transfer matrix of the system for special boundary conditions. That is we show that certain boundary non-local charges commute with the corresponding open transfer matrix.  Note that the derivation of the non-local charges and the symmetry of the open transfer matrix with the aforementioned boundary conditions rely on solely algebraic considerations, and therefore these findings are independent of the choice of representation. In fact, representations of certain boundary non-local charges turn out to commute with the generators of the affine Hecke algebra.  As a consequence the local  Hamiltonian of the open spin chain with a particular choice of boundary conditions, written in terms of the affine Hecke algebra generators, also commutes with the representations of the non-local charges.

A comment is in order on the derivation of the boundary non-local charges. This is in fact the first time that conserved boundary non-local charges are derived explicitly for the $U_{q}(\widehat{gl_{n}})$ case (see also \cite{neg}) with the so called `soliton preserving' (SP) boundary conditions \cite{DVGR, done} emerging from (\ref{re}). From the physical point of view the SP boundary conditions `compel' a particle-like excitation carrying a representation of $U_{q}(gl_{n})$ to reflect to itself ---no multiplet changing occurs. In \cite{dema} on the other hand boundary non-local charges were obtained from the field theory point of view, for `soliton non-preserving' (SNP) boundary conditions \cite{cor}--\cite{doikoupr}. These boundary conditions force an excitation to reflect to its `conjugate', i.e. to reflect to an excitation which carries the conjugate representation. Let us point out that the remarkable feature of the present approach is that allows the investigation of boundary conditions that have not been necessarily treated at the classical level, such as the SP ones which will be discussed subsequently.

\section{The Yang--Baxter equation}

In this section the necessary algebraic background is set  by reviewing basic definitions regarding the Hecke algebras and the affine quantum algebras. We shall briefly recall how solutions of the Yang--Baxter equation arise by exploiting the existence of appropriate representations of the Hecke algebra. Also, representations of (affine) quantum algebras will be obtained by investigating the asymptotic behaviour of the corresponding closed quantum spin chain. 

\subsection{The Hecke algebra}

It will be convenient to rewrite the Yang--Baxter equation (\ref{YBE}) in a slightly modified form by introducing the object \cite{jimbo} \be \check R(\lambda) = {\cal P}\ R(\lambda) \label{ch} \ee  $ {\cal P}$ is the permutation operator acting on  ${\mathbb V}^{ \otimes 2}$: $ ~{\cal  P}( a  \otimes b) = b \otimes a~$ for any vectors $a,~b \in {\mathbb V}$.\\ The $\check R$ matrix satisfies the braid Yang--Baxter equation \be \check R_{12}(\lambda_{1} -\lambda_{2})\ \check R_{23}(\lambda_{1})\ \check R_{12}(\lambda_{2}) =\check R_{23}(\lambda_{2})\ \check R_{12}(\lambda_{1})\ \check R_{23}(\lambda_{1}-\lambda_{2}). \label{ybe2} \ee acting on ${\mathbb V} \otimes{\mathbb V} \otimes {\mathbb V}$, and as usual $~R_{12} = R\otimes {\mathbb I}, ~~R_{23} = {\mathbb I} \otimes R~$.
In what follows we shall recall how solutions to the braid Yang--Baxter equation can be obtained using the Hecke algebra ${\cal H}_{N}(q)$ of type $A_{N-1}$ \cite{hecke, hecke1}.\\
\\
{\bf  Definition 2.1.} {\it The Hecke algebra ${\cal H}_{N}(q)$ is defined by the generators $g_{l}$, $l=1,\ldots ,N-1$ satisfying the following relations:}
\be &&(g_{l}-q)\ (g_{l} +q^{-1}) =0 \non\\ && g_{l}\ g_{l+1}\ g_{l} = g_{l+1}\ g_{l}\ g_{l+1}, \non\\ && [ g_{l},\ g_{m}]  =0, ~~~|l-m| >1.\label{braid} \ee 
 The structural similarity between (\ref{ybe2}) and the second relation of (\ref{braid}) (braid relation) is apparent, it is therefore quite natural to seek for  representations of the Hecke algebra as candidate solutions of the Yang--Baxter equation \cite{jimbo}.\\
For convenience an alternative set of generators of ${\cal H}_{N}(q)$ may be considered i.e. ${\cal U}_{l} = g_{l} -q$, which satisfy the following defining relations
\be && {\cal U}_{l}\ {\cal U}_{l} = \delta\  {\cal U}_{l} \non\\ 
&& {\cal U}_{l}\ {\cal U}_{l+1}\ {\cal U}_{l} - {\cal U}_{l}= {\cal U}_{l+1}\ {\cal U}_{l}\ {\cal U}_{l+1}- {\cal U}_{l+1} \non\\ 
&& [{\cal U}_{l},\ {\cal U}_{m}]=0, ~~~|l-m|>1,  \label{hecke} \ee where $\delta =-(q+q^{-1})$ and $q=e^{i \mu}$. 

\subsection{Solutions of the Yang--Baxter equation} 

As pointed out in \cite{jimbo} tensor product representations of ${\cal H}_{N}(q)$ $\rho: {\cal H}_{N}(q) \to \mbox{End}({\mathbb V}^{\otimes N})$ provide solutions to Yang--Baxter equation, i.e.
\be \check R_{l\ l+1}(\lambda) = \sinh  (\lambda+i\mu)\ {\mathbb I} + \sinh  \lambda\ \rho({\cal U}_{l}). \label{r} \ee It is worth remarking that by virtue of the first of the equations (\ref{hecke}) the unitarity of the $\check R$ matrix, can be verified, i.e. \be \check R(\lambda)\ \check R(-\lambda) \propto{\mathbb I}, ~~~~\lambda \neq \pm i\mu. \label{uni1} \ee
Let  now ${\mathbb V} ={\mathbb C}^{n}$ and  \be (\hat e_{ij})_{kl} =\delta_{ik}\ \delta_{jl} \label{eh} \ee define also the matrix $U$ on $({\mathbb C}^{n})^{\otimes 2}$ of the form \cite{jimbo}: \be U = \sum_{i \neq j =1}^{n}(\hat e_{ij} \otimes \hat e_{ji} - q^{-sgn(i-j)} \hat e_{ii} \otimes \hat e_{jj}). \label{sol} \ee 
Then the following representation is obtained 
$\rho: {\cal H}_{N}(q) \to \mbox{End}(({\mathbb C}^{n})^{\otimes N})$ such that
\be \rho({\cal U}_{l}) = {\mathbb I} \otimes \ldots \otimes U \otimes \ \ldots \otimes {\mathbb I} \label{sol3} \ee acting non-trivially on ${\mathbb V}_{l} \otimes {\mathbb V}_{l+1}$. 
As noted in \cite{jimbo} the latter solution (\ref{r}), (\ref{sol}), (\ref{sol3}) corresponds to the fundamental representation of the $U_{q}(\widehat{gl_n})$. 

From (\ref{ch}) and (\ref{r}) it follows that the $R$ matrix in the fundamental representation can be expressed as 
\be R(\lambda) =  {\cal P}(\sinh (\lambda +i\mu)\ {\mathbb I} + \sinh  \lambda\ U). \label{sol2} \ee It immediately follows from (\ref{uni1}) that the $R$ matrix is also unitary, in particular \be  R(\lambda)\ \hat R(-\lambda) \propto{\mathbb I} ~~~~\mbox{where} ~~~~~\hat R(\lambda) = {\cal P}\ R(\lambda)\ {\cal P}. \label{uni2} \ee
Notice that the lower indices of the $R$ matrix are omitted in (\ref{sol2}) `index free' notation. A useful remark is in order here, note that whenever we write $R$ (`index free' notation) we simply mean that the $R$ matrix acts abstractly on ${\mathbb V}^{\otimes 2}$.  $R_{lm}$ on the other hand acts on specific spaces ${\mathbb V}_{l} \otimes {\mathbb V}_{m}$, where the indices $l$, $m$ denote the position in a tensor product sequence of spaces ${\mathbb V}_{1} \otimes {\mathbb V}_{2} \otimes \ldots \otimes {\mathbb V}_{l} \otimes \ldots \otimes {\mathbb V}_{m} \otimes  \ldots$. In fact a more general remark can be stated: consider the operator ${\cal O}$ (`index free' notation) acting on ${\mathbb V}^{\otimes M}$, then  ${\cal O}_{i_{1} i_{2} \ldots i_{M}}$ acts specifically on ${\mathbb V}_{i_{1}} \otimes {\mathbb V}_{i_{2}} \otimes \ldots \otimes {\mathbb V}_{i_{M}}$, again the indices $~i_{j}, ~j\in \{1, \ldots ,M\}~$ characterize the position in a tensor product sequence.

The $R$ matrix (\ref{sol2}) is written in the homogeneous gradation, it can be however given in the principal gradation by means of a simple gauge transformation (see e.g. \cite{neg}), namely \be R_{12}^{(p)}(\lambda) = {\cal V}_{1}(\lambda)\  R_{12}^{(h)}(\lambda)\ {\cal V}_{1}(-\lambda) \label{gauge} \ee where \be {\cal V}(\lambda)  =diag(1,\ e^{{2\over n}\lambda},\ \ldots, e^{{(n-1)2 \over n}\lambda}). \label{v} \ee It is instructive to write down explicit expressions for the $R$ matrix in both homogeneous and principal gradations: \be &&R^{(h)}(\lambda) = a(\lambda) \sum_{i=1}^{n} \hat e_{ii} \otimes \hat e_{ii}+ b(\lambda) \sum_{i\neq j=1}^{n} \hat e_{ii} \otimes \hat e_{jj} + c \sum_{i\neq j =1}^{n} e^{ -sgn(i-j)\lambda} \hat e_{ij}\otimes \hat e_{ji} \non\\ &&R^{(p)}(\lambda) = a(\lambda) \sum_{i=1}^{n} \hat e_{ii}
  \otimes \hat e_{ii}+ b(\lambda) \sum_{i\neq j=1}^{n} \hat e_{ii} \otimes \hat e_{jj} + c \sum_{i\neq j =1}^{n} e^{((i-j){2 \over n} -sgn(i-j))\lambda} \hat e_{ij}\otimes \hat e_{ji}, \non\\ && a(\lambda)=\sinh (\lambda+i\mu), ~~~b(\lambda) = \sinh \lambda, ~~~c = \sinh i \mu. \label{rhp} \ee The latter $R$ matrices (\ref{rhp}) satisfy in addition to unitarity (\ref{uni2}) the following relations:
\be R_{12}^{t_{1}}(\lambda)\ M_{1}\ R_{12}^{t_{2}}(-\lambda -2i\rho)\ M_{1}^{-1} \propto {\mathbb I}, \ee
where we define $M$ as:
\be  &&  M_{ij}=e^{i\mu(n-2j+1)}\ \delta_{ij}, ~~~\mbox{homogeneous gradation} \non\\ && M_{ij} = \delta_{ij}, ~~~~~~~~~~~~~~~~~\mbox{principal gradation} \non\\ && i,j \in \{1, \ldots ,n \}. \label{M} \ee and \be \Big [M_{1}\ M_{2},\ R_{12}(\lambda) \Big ]=0. \ee

\subsection{The quantum Kac--Moody algebra $U_{q}(\widehat{ sl_n})$}

It will be useful for what is described in the subsequent sections  to recall the basic definitions regarding the affine quantum algebras. Let \be a_{ij} = 2 \delta_{ij} - (\delta_{i\ j+1}+ \delta_{i\ j-1} +\delta_{i1}\ \delta_{jn}+\delta_{in}\ \delta_{j1}), ~~i,\ j\in \{1, \ldots , n \} \ee  be the Cartan matrix of the affine Lie algebra
${\widehat{sl_n}}$\footnote{For the $\widehat{sl_{2}}$ case in particular \be a_{ij} =2\delta_{ij} -2 (\delta_{i1}\ \delta_{j2} +\delta_{i2}\ \delta_{j1}), ~~i,\ j \in \{ 1, 2\}\ee} \cite{kac}. Also define: \begin{eqnarray} && [m]_{q} ={q^{m} -q^{-m} \over q -q^{-1}}, ~~~[m]_{q}!= \prod_{k=1}^{m}\ [k]_{q},~~~[0]_{q}! =1 \non\\  &&\left [ \begin{array}{c}
m \\
n \\ \end{array} \right  ]_{q} = {[m]_{q}! \over [n]_{q}!\ [m-n]_{q}!}, ~~~m>n>0.  \end{eqnarray}\\
\\
{\bf Definition 2.2.} {\it The quantum affine enveloping
algebra $U_{q}(\widehat{ sl_n})$ has the
Chevalley-Serre generators} \cite{jimbo, drinf}  $e_{i}$, $f_{i}$,
$q^{\pm {h_{i}\over 2}}$, $i\in \{1, \ldots, n\}$ {\it obeying the defining relations:} \be
&& \Big [q^{\pm {h_{i}\over 2}},\ q^{\pm {h_{j}\over 2}} \Big]=0\, \qquad q^{{h_{i}\over 2}}\ e_{j}=q^{{1\over
2}a_{ij}}e_{j}\ q^{{h_{i}\over 2}}\, \qquad q^{{h_{i}\over 2}}\ f_{j}
= q^{-{1\over 2}a_{ij}}f_{j}\ q^{{h_{i}\over 2}}, \non\\
&& \Big [e_{i},\ f_{j}\Big ] = \delta_{ij}{q^{h_{i}}-q^{-h_{i}} \over q-q^{-1}},
~~~~i,j \in \{ 1, \ldots,n \}
\label{1} \ee
{\it and the $q$ deformed Serre relations 
\begin{eqnarray} && \sum_{n=0}^{1-a_{ij}} (-1)^{n}
\left [ \begin{array}{c}
  1-a_{ij} \\
   n \\ \end{array} \right  ]_{q} 
\chi_{i}^{1-a_{ij}-n}\ \chi_{j}\ \chi_{i}^{n} =0, ~~~\chi_{i} \in \{e_{i},\ f_{i} \}, ~~~ i \neq j. \label{chev} \end{eqnarray}}
{\it Remark}: The generators $e_{i}$, $f_{i}$, $q^{\pm h_{i}}$ for $i\in \{1, \ldots, n-1\}$ form the $U_{q}( sl_n)$ algebra. Also, $q^{\pm h_{i}}=q^{\pm (\epsilon_{i} -\epsilon_{i+1})}$, where the elements
$q^{\pm \epsilon_{i}}$ belong to $U_{q}(gl_n)$. Recall that $U_{q}(gl_n)$ is derived by adding to $U_{q}( sl_n)$ the elements $q^{\pm \epsilon_{i}}$ $i\in \{1, \ldots, n\}$  so that $q^{\sum_{i=1}^{n}\epsilon_{i}}$ belongs to the center (for more details see \cite{jimbo}).
Furthermore, as noted in \cite{jimbo} there exist the elements ${\cal E}_{ij} \in U_{q}(gl_n)$ $i\neq j$, with ${\cal E}_{i\ i+1} =e_{i}$, ${\cal  E}_{i+1\ i} =f_{i}$ $i\in \{1, \ldots n-1\}$ and \be {\cal E}_{ij} &=& {\cal E}_{ik}\ {\cal E}_{kj}-q^{\mp 1}{\cal E}_{kj}\ {\cal E}_{ik},~~~j \lessgtr k \lessgtr i, ~~~i,j \in \{1, \ldots ,n  \}. \label{ee} \ee It is clear that  ${\cal E}_{ij} \in U_{q}(gl_n)$ because they can be written solely in terms of the generators $e_{i}$, $f_{i}$, $~i\in \{1, \ldots, n-1 \}$.  $\square$

The algebra ${\cal A} = U_{q}(\widehat{gl_{n}})$ is equipped with a coproduct 
$\Delta: {\cal A}
\to {\cal A} \otimes {\cal A}$ such that
\be \Delta(e_{i}) = q^{- {h_{i} \over 2}} \otimes y + y \otimes q^{{h_{i} \over 2}}, ~~y \in \{e_{i},\ f_{i} \},
\qquad \Delta(q^{\pm{\epsilon_{i} \over 2}}) = q^{\pm{\epsilon_{i} \over 2}} \otimes
q^{\pm{\epsilon_{i} \over 2}} .\label{cop} \ee  It will be useful to define also $\Delta': {\cal A} \to {\cal A} \otimes {\cal A}$. Let $\Pi$ be the `shift operator'   $~\Pi:\ {\cal U}_{1} \otimes {\cal U}_{2}\ \to\  {\cal U}_{2}  \otimes {\cal U}_{1}$   then  \be \Delta'(x)\   = \Pi \circ \Delta(x) ,~~~x \in {\cal A}. \label{perm} \ee The $L$-fold coproduct may be derived by using the recursion relations \be \Delta^{(L)} = (\mbox{id} \otimes \Delta^{(L-1)}) \Delta , ~~~~~~\Delta^{'(L)} = (\mbox{id} \otimes \Delta^{(L-1)}) \Delta' . \label{cop2} \ee  As customary, $\Delta^{(2)} = \Delta$ and  $\Delta^{(1)} = \mbox{id}$. Finally by using (\ref{cop2}) we may derive explicit expressions for any $L$ as: \be &&\Delta^{(L)}(y) = \sum_{l=1}^{L} q^{-{h_{i} \over 2}}\otimes \ldots \otimes q^{-{h_{i} \over 2}} \otimes \underbrace{y}_{\mbox{ $l$ position}}
\otimes q^{{h_{i} \over 2}} \otimes \ldots \otimes q^{{h_{i} \over 2}}, ~~~y \in \{e_{i},\ f_{i} \} \non\\  &&\Delta^{(L)}(q^{\pm {\epsilon_{i} \over 2}}) = q^{\pm {\epsilon_{i} \over 2}}\otimes \ldots
\otimes q^{\pm {\epsilon_{i} \over 2}}. \label{ncop} \ee The opposite coproduct $\Delta^{(L)op}(x)$ can be also derived from $\Delta^{(L)}(x)$, $x \in {\cal A}$
(\ref{ncop}) by $q^{\pm {h_{i} \over 2}} \to q^{\mp {h_{i} \over 2}}$.  We should note that for
general $L$  $\Delta^{(L)op}(x) \neq \Delta^{'(L)}(x)$, and they only coincide when $L=2$. 
Note that for $e_{i}$, $f_{i}$, $i\in \{1, \ldots, n-1 \}$ the coproducts are restricted to the non-affine case.

\subsection{Evaluation representation and Lax operators} 

It will be useful for the following to consider the universal ${\cal R}$  matrix, which is a solution of the
universal Yang--Baxter equation,
\be {\cal R}_{12}\ {\cal R}_{13}\ {\cal R}_{23} ={\cal R}_{23}\ 
{\cal R}_{13}\ {\cal R}_{12}. \label{uni} \ee
\\
$\blacklozenge$ {\it Homogeneous gradation:} Let us now consider the evaluation representation \cite{jimbo}
$\pi_{\lambda}:{\cal A} \to \mbox{End}({\mathbb C}^n)$ defined 
as\footnote{Also, \be \pi_{\lambda}
(q^{{\epsilon_{i} \over 2}}) = q^{{\hat e_{ii} \over 2}} \label{eval2} \ee 
}:
\be &&\pi_{\lambda}(e_{i})=\hat e_{i\ i+1}, ~~\pi_{\lambda}(f_{i})=\hat 
e_{i+1\ i},~~~
\pi_{\lambda}(q^{{h_{i}\over 2}}) = q^{{1\over 2}(\hat e_{ii} -\hat 
e_{i+1\ i+1})}, ~~
i=1,\ldots, n-1 \non\\ &&\pi_{\lambda}(e_{n})=e^{-2 \lambda}\hat e_{n 1}, ~~\pi_{\lambda}
(f_{n})=e^{2 \lambda}\hat e_{1 n},~~~ \pi_{\lambda}(q^{{h_{n}\over 2}})= q^{{1\over 2}
(\hat e_{nn} -\hat e_{1 1})}. \label{eval} \ee
It follows from (\ref{ee}), (\ref{q1}) and (\ref{eval}) that \be \pi_{0}({\cal E}_{ij})= \pi_{0}(\hat
{\cal E}_{ij})= \hat e_{ij} ~~~i,~ j \in \{1, \ldots, n\}. \label{evale} \ee
Then define the Lax operator \be {\cal L}(\lambda) = (\pi_{\lambda} \otimes \mbox{id}) {\cal R} ~~~~
\mbox{also}  ~~~~~R(\lambda_{1} -\lambda_{2}) = (\pi_{\lambda_{1}} \otimes \pi_{\lambda_{2}})  {\cal R}.
\label{def1} \ee $R \in \mbox{End}(({\mathbb C}^{n})^{\otimes 2})$ satisfies apparently the Yang--Baxter
 equation (\ref{YBE}), while ${\cal L}  \in \mbox{End}({\mathbb C}^{n}) \otimes {\cal A}$ satisfies
a fundamental algebraic relation, which is immediate consequence of (\ref{uni}) and (\ref{def1})
\be  R_{ab}(\lambda_{1} -\lambda_{2})\ {\cal L}_{a}(\lambda_{1})\  {\cal L}_{b}(\lambda_{2}) =
{\cal L}_{b}(\lambda_{2})\  {\cal L}_{a}(\lambda_{1})\  R_{ab}(\lambda_{1} -\lambda_{2}). \label{funda} \ee
Note that the algebra defined by (\ref{funda}) is endowed with a coproduct $\Delta: {\cal A} \to {\cal A} \otimes {\cal A}$ i.e. \be
(\mbox{id} \otimes \Delta){\cal L}(\lambda)  = {\cal L}_{13}(\lambda) \ {\cal L}_{12}(\lambda)\ \to\  {\cal L}_{ij}(\lambda)  = \sum_{k=1}^{n} {\cal L}_{kj}(\lambda)  \otimes {\cal L}_{ik}(\lambda),~~~~~
 i,\ j \in \{1,\ldots, n \} . \label{cob1} \ee  A solution of the above equation (\ref{funda}) may be written in
the simple form below (see also \cite{ft11}--\cite{moras}) \be
{\cal L}(\lambda)= e^{\lambda} {\cal L}^{+} - e^{-\lambda} {\cal
L}^{-}, \label{lh0} \ee with the matrices ${\cal L}^{+}$, ${\cal
L}^{-}$ being upper (lower) triangular, i.e. \be {\cal L}^{+} =
\sum_{i\leq j =1}^{n} \hat e_{ij} \otimes t_{ij}, ~~~~~{\cal
L}^{-} = \sum_{i \geq j =1}^{n} \hat e_{ij} \otimes
t_{ij}^{-}.\label{lh} \ee $t_{ij},~t_{ij}^{-} \in U_{q}(gl_{n})$
are defined explicitly in appendix A, and they also form simple coproducts as shown in appendix A (\ref{ns2}).
It can be verified by inspection that the Lax operator and the $R$ matrix satisfy the following linear intertwining relations (see e.g.
\cite{jimbo, jimbo2, leclair}) \be &&  (\pi_{\lambda}\otimes \mbox{id})
\Delta'(x)\ {\cal L}(\lambda) ={\cal L}(\lambda)\ (\pi_{\lambda}\otimes
\mbox{id})\Delta(x),  \non\\ && (\pi_{\lambda}\otimes \pi_{0})\Delta'(x)\
R(\lambda) =R(\lambda)\ (\pi_{\lambda}\otimes \pi_{0})\Delta(x), ~~~x \in {\cal A}.
\label{inter} \ee Actually, once having verified such relations for the known $R$ matrix generalized intertwining relations (\ref{inter})
can be derived, and the Lax operator ${\cal L}$ may be deduced.\\
\\
$\blacklozenge$ {\it Principal gradation:} The evaluation representation in the principal gradation can be obtained by virtue of the gauge transformation \be \tilde \pi_{\lambda}(x) = {\cal V}(\lambda)\ \pi_{\lambda}(x)\ {\cal V}(-\lambda), ~~~x  \in {\cal A} \label{vv} \ee where ${\cal V}(\lambda)$ is given by (\ref{v}). Then we can write $\tilde \pi_{\lambda}:{\cal A} \to \mbox{End}({\mathbb C}^n)$ such that 
\be &&\tilde \pi_{\lambda}(e_{i})=e^{-{2 \lambda \over n}}\hat e_{i\ i+1}, ~~\tilde \pi_{\lambda}(f_{i})=e^{{2 \lambda \over n}}\hat e_{i+1\ i},~~~ \tilde \pi_{\lambda}(q^{{h_{i}\over 2}}) = q^{{1\over 2}(\hat e_{ii} -\hat e_{i+1\ i+1})}, ~~i=1,\ldots, n-1 \non\\ &&\tilde \pi_{\lambda}(e_{n})=e^{-{2  \lambda \over n}}\hat e_{n 1}, ~~\tilde \pi_{\lambda}(f_{n})=e^{{2 \lambda \over n}}\hat e_{1 n},~~~ \tilde \pi_{\lambda}(q^{{h_{n}\over 2}})= q^{{1\over 2}(\hat e_{nn} -\hat e_{1 1})}. \label{eval'} \ee Note that $\pi_{0} =\tilde \pi_{0}$. The ${\cal L}$ matrix in the principal gradation takes then the following form \be {\cal L}(\lambda) &=& \sum_{i=1}^{n} \hat e_{ii} \otimes 2 \sinh (\lambda +i\mu \epsilon_{i}) +\sum_{i<j } e^{((i-j){2\over n} +1)\lambda }(1-\delta_{i1} \delta_{jn})\hat e_{ij} \otimes t_{ij}+e^{\lambda -{2\over n}\lambda} \hat e_{n1} \otimes t_{n1}^{0} \non\\ &-& \sum_{i>j} e^{((i-j){2\over n} -1)\lambda }(1-\delta_{in} \delta_{j1})\hat e_{ij} \otimes t_{ij}^{-} - e^{-\lambda +{2\over n}\lambda} \hat e_{1n} \otimes t_{1n}^{0-}. \label{lp} \ee The elements $t_{1n}^{0-}$ and $t_{n1}^0$ are associated to the affine generators (see appendix A). Intertwining relations as in (\ref{inter}) for the Lax operator ${\cal L}$ in the principal gradation also hold with $\pi \to \tilde \pi$.

Finally it can be directly shown \cite{jimbo} that $\rho({\cal U}_{l})$, and via (\ref{r}) the $\check R$ matrix, commute with the $U_{q}(gl_{n})$ generators. Indeed, it is straightforward to show for $N=2$ that \be \Big [ \rho({\cal U}_{1}),\ \pi_{0}^{\otimes 2}(\Delta(x)) \Big ] = \Big [ \check R_{12}(\lambda),\ \pi_{0}^{\otimes 2}(\Delta(x)) \Big ] =  0, ~~~x\in U_{q}(gl_{n}). \label{comte} \ee Then from (\ref{ncop}) and $\rho({\cal U}_{l})$ (\ref{sol}), (\ref{sol3}) it follows for any $N$ \be\Big [ \rho({\cal U}_{l}),\ \pi_{0}^{\otimes N}(\Delta^{(N)}(x)) \Big ] = \Big [ \check R_{l\ l+1}(\lambda),\ \pi_{0}^{\otimes N}(\Delta^{(N)}(x)) \Big ] =  0, ~~~x\in U_{q}(gl_{n}), ~~~l\in \{1, \ldots N-1 \}. \label{comten} \ee The latter relations will turn out to be instrumental in the study of the open spin chain Hamiltonian as will be clear in section 4.

\subsection{The closed spin chain}

Tensor products of the quantum algebra may be now considered yielding the $N$ site closed spin chain. This construction is achieved by using the Quantum Inverse Scattering Method (QISM) an approach initiated in \cite{FT, korepin, fad}, providing also one of the main motivations for the formulation and study of quantum groups \cite{jimbo, drinf, tak, ft11}.

The first step is to introduce the monodromy matrix $T(\lambda) \in \mbox{End}({\mathbb C}^{n}) \otimes {\cal A}^{\otimes N}$ being also a solution of (\ref{funda}) and defined as
 \be T_{0}(\lambda) = (\mbox{id} \otimes \Delta^{(N)}){\cal L}(\lambda) = {\cal L}_{0N}(\lambda)\ {\cal L}_{0\ N-1}(\lambda) \ldots {\cal L}_{02}(\lambda)\ {\cal L}_{01}(\lambda). \label{mono} \ee  The notation ${\cal L}_{0l}$ implies that ${\cal L}$ acts on $\underbrace{{\mathbb C}^{n}}_{\mbox{ $0$ position}} \otimes \ldots \otimes \underbrace{{\cal A}}_{\mbox{ $l$ position}} \otimes \ldots$, where the numbering in the tensor product sequence is considered from $0$ to $N$. Traditionally the indices $l \in \{1, \ldots, N \}$, associated to the so called `quantum' spaces, are suppressed from $T$, and we only keep the index $0$ corresponding to the so called `auxiliary' space. Here, the homogeneous monodromy matrix is considered for simplicity, although we could have added inhomogeneities $\Theta_{i}$ at each site of the spin chain (see e.g. \cite{deve}). We could have also chosen a different representation for the auxiliary space, but in order to keep things uncomplicated we consider it to be $n$ dimensional (fundamental representation). Such a choice simplifies dramatically the subsequent computations without however reducing their generality.

The transfer matrix of the system, is simply defined as the trace over the `auxiliary' space, i.e. \be t(\lambda) =tr_{0}\ T_{0}(\lambda) \label{transfer} \ee thus $t(\lambda)$ is an element of ${\cal A}^{\otimes N}$. It can be then shown via (\ref{funda}) that  \cite{FT}  \be \Big [ t(\lambda),\ t(\lambda')\Big] =0, \label{com} \ee a condition that ensures the integrability of the model. As no representation has been specified on the `quantum' spaces, our description is purely algebraic at this stage, that is the entailed results are independent of the choice of representation, and thus they are universal.  It is not until we assign particular representations on the `quantum' spaces that the algebraic construction (\ref{mono}), (\ref{transfer}), consisting of $N$ copies of ${\cal A}$, acquires a physical meaning as a quantum spin chain. Once having specified the representations of the quantum spaces one may diagonalize the transfer matrix (\ref{transfer}), which is the quantity that encodes all the physical information of the system, and derive the corresponding Bethe ansatz equations \cite{FT}.

Generalized intertwining relations may be obtained by induction from (\ref{inter}) for the monodromy matrix (\ref{mono}) as well (see also \cite{kor, doikou}),
\be && (\pi_{\lambda}\otimes \mbox{id}^{\otimes N})\Delta^{'(N+1)}(x)\ T(\lambda) =T(\lambda)\ (\pi_{\lambda}\otimes \mbox{id}^{\otimes N}) \Delta^{(N+1)}(x), ~~~x\in {\cal A},  \label{intert} \ee  $\Delta^{(N+1)}$ is treated as a two site coproduct in such a way that on the first site the representation $\pi_{\lambda}$ acts, while on the `second' site ---which is a composite of $N$ sites--- $\mbox{id}^{\otimes N}$ acts (no specific choice of representation for the quantum spaces) see also (\ref{cop2}). The intertwining relations (\ref{intert}) yield important commutation relations between the generators of ${\cal A}$ and the entries of the monodromy matrix, enabling the investigation of the symmetry of the transfer matrix \cite{kor, doikou}, as will become clear later. For the closed spin chain, for generic values of $q$ no symmetry has been identified for the transfer matrix. It is actually the open spin chain --which will be discussed later-- with special boundary conditions that enjoys the full  $U_{q}(gl_n)$ symmetry \cite{pasa, kusk, menes}. The symmetry of the closed spin chain  has been only derived for the case where $q$ is root of unity \cite{kor}, \cite{mac}--\cite{deg}. In particular, as argued in \cite{kor}, \cite{mac}--\cite{deg} the periodic or twisted spin chain, for $q$ root of unity,  enjoys the $sl_n$ loop algebra symmetry.

We should finally mention that intertwining relations of the type (\ref{intert}) for the monodromy matrix in the principal gradation may be deduced by acting on (\ref{intert}) with the gauge transformation ${\cal V}_{0}(\lambda)$ and using also (\ref{gauge}). In particular, the monodromy matrix satisfies the same form of relations (\ref{intert}) but with ($\pi_{\lambda} \to \tilde \pi_{\lambda}$).

\subsection{Non-local charges}

The asymptotic behaviour of the monodromy matrix will be investigated for both the homogeneous and principal gradation. The reason why such an investigation is of great relevance is because it bears, as will become clear, coproducts of ${\cal A}$ (see also e.g. \cite{tak, ft11}).\\ 
\\
$\blacklozenge$ {\it Homogeneous gradation:} 
We shall first examine the asymptotic behaviour of the monodromy matrix in the homogeneous gradation, which provides coproducts of $U_{q}(gl_n)$.
In the following the asymptotic forms of ${\cal L}$ and $T$ are treated as $n \times n$ matrices with entries being elements of $U_{q}(gl_n)$, $U_{q}(gl_n)^{\otimes N}$ respectively.

The ${\cal L}$ matrix (\ref{lh0}), (\ref{lh}) as $ \lambda \to \pm \infty$ reduces to an upper (lower) triangular form (recall that $q=e^{i\mu}$, also for the asymptotics we consider $\mu$ to be finite)
\be  {\cal L}(\lambda \to \pm \infty) \propto{\cal L}^{\pm} \label{asy} \ee
where indeed the matrices ${\cal L}^{+}$ (${\cal L}^{-}$) are upper (lower) triangular matrices, with entries being elements of $U_{q}(gl_{n})$ and given in 
(\ref{lh}). It is then straightforward from (\ref{mono}) and (\ref{asy}) to write down the asymptotic behaviour of the monodromy matrix as $ \lambda\to \pm \infty$, i.e. (here for simplicity the `auxiliary' space index $0$ is suppressed from $T$ (\ref{mono}))
\be T(\lambda\to \pm \infty) \propto  T^{\pm(N)}. \label{+1}\ee As expected the matrices $T^{\pm(N)}$ are upper (lower) triangular with the non-zero entries being elements of $U_{q}(gl_{n})^{\otimes N}$:  
\be T_{ij}^{+(N)} = \Delta^{(N)}(t_{ij}) ~~i \leq j, ~~~~~ T_{ij}^{-(N)} = \Delta^{(N)}(t^{-}_{ij}) ~~i \geq j, ~~~i,j \in \{1, \ldots, n \} . \label{reppr} \ee Let us stress that all the above expressions are independent of $\lambda$, and the coproducts appearing in (\ref{reppr}) are restricted to the non-affine case, providing  tensor product realizations of $U_{q}(gl_n)$. The algebraic objects $T^{+(N)}$, $T^{-(N)}$ provide actually coproducts of the upper lower Borel subalgebras respectively.\\ 
\\
$\blacklozenge$ {\it  Principal gradation:} 
In order to extract expressions associated to the affine generators of ${\cal A}$ it is convenient to consider the asymptotic expansion of the monodromy matrix in the principal gradation. In this case we keep in the $T(\lambda \to \infty)$ expansion zero order terms and $e^{\pm {2\over n} \lambda}$ terms as well \cite{doikou}. The asymptotic behaviour of  ${\cal L}$ in the principal gradation (\ref{lp}) as $\lambda\to \pm \infty$ is given by (see also \cite{dema}), \be {\cal L}( \lambda\to \pm \infty) \propto ( D^{\pm} +e^{\mp{2\over n}  \lambda}\ B^{\pm}+\ldots), \label{asyp}  \ee
where the non-zero entries of the $D^{\pm}$, $B^{\pm}$ matrices are elements of ${\cal A}$, 
\be && D^{\pm}_{ii} = t_{ii}^{\pm 1}, ~~~i \in \{ 1, \ldots, n\}, ~~~~~B^{+}_{n1} = t_{n1}^{0}, ~~~~~B^{-}_{1n} =t_{1n}^{0-}, \non\\ && B^{+}_{i\ i+1} = t_{i\ i+1}, ~~~~~B^{-}_{i+1\ i} = t_{i+1\ i}^{-}, ~~~~i\in \{1,\ldots , n-1\}. \label{db} \ee
Consequently, the asymptotics of the monodromy matrix take the form \be T(\lambda\to \pm \infty)\propto (D^{\pm(N)} +e^{\mp {2\over n} \lambda} B^{\pm(N)}+\ldots) \label{astt2} \ee with the non-zero entries of $ D^{\pm(N)}$ and $ B^{\pm(N)}$ being elements of ${\cal A}^{\otimes N}$  given by: \be && D^{\pm(N)}_{ii} = \Delta^{(N)}(t_{ii}^{\pm 1}),  ~~~~~B_{n1}^{+(N)} = \Delta^{(N)}(t_{n1}^0), ~~~~~B^{-(N)}_{1 n } = \Delta^{(N)}(t_{1n}^{0-}) \non\\ &&B_{i\ i+1}^{+(N)} =\Delta^{(N)}(t_{i\ i+1}), ~~~~~B^{-(N)}_{i+1\ i } = \Delta^{(N)}(t^{-}_{i+1\ i}). \label{astt3} \ee Recall that $t_{ij}$, $t_{ij}^{-}$, $t_{n1}^0$, $t_{1n}^{0-}$ and their coproducts are defined in appendix A. 

The lowest orders of the monodromy matrix asymptotics  in both homogeneous and principal gradation gave rise, as expected, to coproduct realizations of ${\cal A}$. The entailed quantities (\ref{reppr}), (\ref{astt3}) are the non-local charges, which for the periodic case for generic values of $q$ do not commute with the transfer matrix of the system.
Our intention is to generalize the process described so far, when open integrable boundaries are implemented. In this case it turns out, as we shall see in the subsequent sections, that some of the induced non-local charges are conserved quantities.

\section{The reflection equation}

After the brief review on the bulk case we may now present the main results regarding the boundary case. In particular, solutions of the reflection equation for the $U_{q}(\widehat{gl_{n}})$ case will be obtained by means of the affine Hecke algebra \cite{doma, male, masa}. Once having available solutions of the reflection equation we shall be able to construct the corresponding open spin chain, and extract the boundary non-local charges along the lines described in \cite{dema, doikou}. Exploiting the existence of intertwining relations analogous to (\ref{inter}), (\ref{intert}) we shall show that certain boundary non-local charges are conserved quantities, that is they commute with the transfer matrix of the open spin chain. The approach that is discussed in the following may be thought of as the boundary analogue of the bulk case described in the previous sections. 

\subsection{The affine Hecke algebra}

As in the case of the Yang--Baxter equation it will be convenient to rewrite the reflection equation (\ref{re}) in a modified form i.e. \be \check R_{12}(\lambda_{1} -\lambda_{2})\ K_{1}(\lambda_{1})\ \check R_{12}(\lambda_{1} +\lambda_{2})\ K_{1}(\lambda_{2})=K_{1}(\lambda_{2})\ \check R_{12}(\lambda_{1} +\lambda_{2})\ K_{1}(\lambda_{1})\ \check R_{12}(\lambda_{1} -\lambda_{2}) \label{re2} \ee acting on ${\mathbb V}\otimes {\mathbb V}$, and as customary $~K_{1} = K \otimes {\mathbb I}$, $~ K_{2} = {\mathbb I} \otimes   K$. The main objective now is the derivation of solutions of (\ref{re2}) by employing representations of the affine Hecke algebra \cite{ahecke}.\\
\\
{\bf Definition 3.1.} {\it The affine Hecke algebra ${\cal H}_{N}^{0}(q,Q)$ is defined by generators $g_{l}$, $l \in \{ 1, \ldots , N-1 \}$ satisfying the Hecke relations (\ref{braid}) and $g_{0}$ obeying:
\be && g_{1}\ g_{0}\ g_{1}\ g_{0} = g_{0}\ g_{1}\ g_{0}\ g_{1}, \non\\ && \Big [g_{0},\ g_{l} \Big] =0, ~~l>1. \label{braid2} \ee} 
The algebra (\ref{braid2}) is apparently an extension of the Hecke algebra defined in (\ref{braid}).
Notice again the structural similarity between (\ref{re2}) and the first relation of (\ref{braid2}), which suggests that representations of  ${\cal H}_{N}^{0}(q,Q)$  should provide candidate solutions of the reflection equation \cite{male}. However, the affine Hecke algebra is rather `big' to be physical, it is thus quite natural to restrict our attention to quotients of ${\cal H}_{N}^{0}(q,Q)$ \cite{doma, male}.\\
\\
{\bf Definition 3.2.} {\it  A quotient of the affine Hecke algebra, called the $B$-type Hecke algebra ${\cal B}_{N}(q,Q)$, is obtained by imposing in addition to (\ref{braid}), (\ref{braid2}) an extra constraint on $g_{0}$, namely \be (g_{0} -Q)(g_{0} +Q^{-1}) =0. \label{quotient} \ee} We can again choose an alternative set of generators of the $B$-type Hecke algebra ${\cal U}_{l}=g_{l}-q$ and ${\cal U}_{0} =g_{0} -Q$ where ${\cal U}_{l}$ satisfy relations (\ref{hecke}) and \be && {\cal U}_{0}\ {\cal U}_{0} = \delta_{0}\ {\cal U}_{0} \non\\   && {\cal U}_{1}\ {\cal U}_{0}\ {\cal U}_{1}\ {\cal U}_{0} - \kappa\ {\cal U}_{1}\ {\cal U}_{0} = {\cal U}_{0}\ {\cal U}_{1}\ {\cal U}_{0}\ {\cal U}_{1}\ -\kappa\ {\cal U}_{0}\ {\cal U}_{1} \non\\ 
&& \Big [{\cal U}_{0},\ {\cal U}_{l} \Big ] =0, ~~~l>1 \label{hecke0} \ee $\delta_{0} =-(Q+Q^{-1})$ and $\kappa = q Q^{-1} +q^{-1} Q$\footnote{The form of the constants $\delta_{0}$ and $\kappa$ follow from the choice ${\cal U}_{l} =g_{l}-q$, ${\cal U}_{0} =g_{0}-Q$ and from the first relation of (\ref{braid}) and also (\ref{braid2}).}. We are free to renormalize ${\cal U}_{0}$ and consequently  $\delta_{0}$ and $\kappa$ (but still  ${\delta_{0}\over \kappa} = -{Q+Q^{-1} \over qQ^{-1} +q^{-1}Q}$).

We shall be mainly interested in representations of a quotient of the $B$-type Hecke algebra denoted as ${\cal CB}_{N}$ and defined by relations (\ref{hecke}), (\ref{hecke0}) and the additional constraint,
\be {\cal U}_{1}\  {\cal U}_{0}\  {\cal U}_{1}\ {\cal U}_{0}=  \kappa\ {\cal U}_{1}\ {\cal U}_{0} ~~~\mbox{or equivalently} ~~~{\cal U}_{0}\  {\cal U}_{1}\  {\cal U}_{0}\ {\cal U}_{1}=  \kappa\ {\cal U}_{0}\ {\cal U}_{1}.  \label{heckeb} \ee 

\subsection{Solutions of the reflection equation} 

It was shown in \cite{doma, male} that tensor representations of quotients of the affine Hecke algebra provide solutions to the reflection equation. For our purposes here we shall make use of the following:\\
\\
{\bf Proposition 3.1.} {\it Tensor representations of ${\cal H}_{N}(q)$ that extend to ${\cal B}_{N}(q,Q)$, $\rho: {\cal B}_{N}(q, Q) \to \mbox{End}({\mathbb V}^{\otimes N})$ provide solutions to the reflection equation, i.e. \be K (\lambda) = x(\lambda) {\mathbb I} +y(\lambda) \rho({\cal U}_{0}), \label{ansatz} \ee with \be x(\lambda)= -\delta_{0}\cosh (2\lambda +i \mu) -
\kappa \cosh 2 \lambda -\cosh 2 i\mu \zeta
\hspace{1cm}
y(\lambda)=2 \sinh 2\lambda\  \sinh i\mu. \label{ansatz2} \ee} {\it Proof}: This may be shown along the lines described in \cite{doma, male}. In particular, the values of $x(\lambda)$ and $y(\lambda)$ can be found by straightforward computation, by substituting the ansatz (\ref{ansatz}) in (\ref{re2}) and also using equations (\ref{hecke}), (\ref{hecke0}). Also, $\zeta$ in (\ref{ansatz2}) is an arbitrary constant. \finproof 

In analogy to the bulk case (\ref{uni1}), (\ref{uni2}) we should mention that by means of the extra constraint (\ref{quotient}) (equivalently the first of relations (\ref{hecke0})) the unitarity of the solutions (\ref{ansatz}) can be shown, i.e. \be K(\lambda)\ K(-\lambda) \propto {\mathbb I}, ~~~~x(\lambda) \neq 0. \ee
We shall hereafter focus on specific representations of ${\cal CB}_{N}$. In particular, we shall introduce a new representation of ${\cal CB}_{N}$. Let $U_{0}$ be a $n \times n$ matrix given by \be U_{0}= -Q^{-1} \hat e_{11} -Q \hat e_{nn} +\hat e_{1n}+ \hat e_{n1}. \label{solb} \ee Then define the matrix ${\cal M}$ on $\mbox{End}(({\mathbb C}^{n})^{\otimes N})$:
\be {\cal M} = {1\over 2 i \sinh i\mu} U_{0} \otimes {\mathbb I} \ldots \otimes {\mathbb I} \label{solb1} \ee acting non-trivially on ${\mathbb V}_{1}$ (${\mathbb V} = {\mathbb C}^{n}$).\\
\\
{\bf Proposition 3.2.}  {\it There exists a representation $\rho:{\cal CB}_{N} \to \mbox{End}(({\mathbb C}^{n})^{\otimes N})$ such that $\rho({\cal U}_{l})$ are defined by (\ref{sol3}), and $\rho({\cal U}_{0}) ={\cal M}$}.\\
{\it Proof:} It can be proved by direct computation using the property $~~\hat e_{ij}\ \hat e_{kl} = \delta_{jk} \hat e_{il}, ~~$ that the matrices $\rho({\cal U}_{l})$ (\ref{sol}), (\ref{sol3}) and $\rho({\cal U}_{0})$ (\ref{solb}), (\ref{solb1}) satisfy relations (\ref{hecke}), (\ref{hecke0}), (\ref{heckeb}), and hence they indeed provide a representation of ${\cal CB}_{N}$ . \finproof

Note that in \cite{masa} an analogous representation of the blob algebra generators is presented. Also in \cite{male} diagonal representations of quotients of the affine Hecke algebra are considered, and they can be attained from our representation (\ref{solb}), (\ref{solb1}) in the limit $Q^{-1} \to \infty$.\\
For the representation of Proposition 3.2 in particular it is convenient to set $Q=i e^{i \mu m}$ then we obtain \be \delta_{0} = -{\sinh i\mu m \over \sinh i\mu}, ~~~~~\kappa = {\sinh i\mu(m-1) \over \sinh i\mu}. \label{numbers} \ee
Let us finally write the entries of the $n \times n$ $K$ matrix using (\ref{ansatz}), Proposition 3.2 and (\ref{numbers}):
\be &&K_{11}(\lambda) = e^{2 \lambda} \cosh i\mu m - \cosh 2 i \mu \zeta, ~~~K_{nn}(\lambda)=e^{-2 \lambda} \cosh i\mu m - \cosh 2i \mu \zeta \non\\ &&K_{1n}(\lambda)=K_{n1}(\lambda)=-i\sinh 2 \lambda, ~~~K_{jj}(\lambda) = \cosh (2\lambda +i m \mu) - \cosh 2i \mu \zeta, \non\\ && j \in \{2, \ldots ,n-1 \}.\label{k}\ee
The $K$ matrix (\ref{k}) is written in the homogeneous gradation, however one can easily obtain the $K$ matrix in the principal gradation via the gauge  transformation
\be K^{(p)}(\lambda) = {\cal V}(\lambda)\ K^{(h)}(\lambda)\ {\cal V}(\lambda) \label{vk} \ee recall ${\cal V}(\lambda)$ is given in (\ref{v}). Then we can write explicitly: 
\be &&K^{(p)}_{11}(\lambda)= K_{11}(\lambda), ~~~K^{(p)}_{nn}(\lambda)= e^{2(n-1){2 \over n} \lambda} K_{nn}(\lambda), \non\\ &&K^{(p)}_{1n}(\lambda)= K^{(p)}_{n1}(\lambda) =e^{(n-1){2 \lambda \over n}} K_{1n}(\lambda),~~~K^{(p)}_{jj}(\lambda)= e^{(2j-2){2 \over n} \lambda}K_{jj}(\lambda) \non\\ && j \in \{2, \ldots ,n-1 \}, \label{kp}\ee 
with $K_{ij}(\lambda)$ given by (\ref{k}).  

It is instructive to compare our solution (\ref{kp}) with the solution found in  \cite{abad} for the $U_{q}(\widehat{gl_{n}})$ case in the principal gradation, which reads as: \be && K^{(AR)}_{11}(\lambda)=  \rho_{a+}\sinh (\epsilon_{+} -{n \theta \over 2}), ~~~ K^{(AR)}_{nn}(\lambda)=  \rho_{a+} e^{(n-2)\theta}\sinh (\epsilon_{+} +{ n \theta \over 2}) \non\\&&  K^{(AR)}_{1n}(\lambda)=\rho_{d+} e^{({n\over 2} -1) \theta}\sinh n\theta, ~~~K^{(AR)}_{n1}(\lambda)=\rho_{c+} e^{({n\over 2} -1) \theta}\sinh n\theta \non\\ && K^{(AR)}_{jj}(\lambda)=  \rho_{a+} e^{(2j-2-n)\theta} \sinh (\epsilon_{+} +{n \theta \over 2}) + \rho_{b+} e^{(2j-2-{n\over 2})\theta} \sinh n\theta, ~~~j\in \{2, \ldots , n-1 \} \label{ar} \ee and the constants appearing in (\ref{ar}) satisfy \be \rho_{c+} \rho_{d+} = \rho_{b+} (\rho_{b+} + \rho_{a+}e^{-\epsilon_{+}}). \label{rest}  \ee We can consider $\rho_{c+} = \rho_{d+}$ without loss of generality (see also \cite{neg}).
Let  ${n\theta \over 2}= \lambda$, and also  multiply (\ref{kp}) by $i$, and (\ref{ar}) by ${e^{{n\theta \over 2} } \over \rho_{c+}}$ (note that we are free to multiply the $K$ matrix by any factor, because it is derived up to an overall multiplication factor anyway).
Then the expression for the $K$ matrix in the principal gradation (\ref{kp}) coincides with (\ref{ar}) provided that the following identifications hold (see also \cite{doikou}):
\be  e^{-\epsilon_{+}}{\rho_{a+} \over \rho_{c+}}  =-2 i\cosh i\mu m,~~~~ e^{\epsilon_{+}} {\rho_{a+} \over \rho_{c+}} =-2 i\cosh 2i \mu \zeta \label{iden} ~~~~\mbox{and} ~~~~{\rho_{b+} \over \rho_{c+}}= i e^{i\mu m}. \label{rest2} \ee The latter identifications (\ref{rest2}) are compatible with the restrictions imposed upon the constants (\ref{rest}) appearing in the solution (\ref{ar}). Note that the solution we find for $n=2$, in the principle gradation (\ref{kp}), coincides with the two dimensional solution found for the $U_{q}(\widehat{sl_{2}})$ case \cite{DVGR, GZ}.

A comment is in order regarding the solutions of the reflection equation (\ref{re}). It is possible to acquire other solutions of the reflection equation (\ref{re}) for the $U_{q}(\widehat{gl_{n}})$ case, using different representations of quotients of the affine Hecke algebra. Such an exhaustive analysis although of great relevance is not attempted here, but it will be undertaken in detail elsewhere. In this work our main aim is to simply illustrate the Hecke algebraic method as a systematic means for solving the reflection equation for systems associated to higher rank algebras, and also provide at least one non-trivial representation of the $B$-type Hecke algebra (\ref{solb}), (\ref{solb1}).

\subsection{The reflection algebra and the open spin chain}

Having at our disposal c-number solutions of (\ref{re}) we may build the more general form of solution of (\ref{re}) as argued in \cite{sklyanin}. To do so it is necessary to define the following object \be \hat {\cal L}(\lambda) = {\cal L}^{-1}(-\lambda). \ee Although the expression for ${\cal L}^{-1}$ is quite intricate we shall only need for the following the asymptotic behaviour as $\lambda \to \infty$. 

The more general solution of (\ref{re}) is then  given by \cite{sklyanin}:
\be {\mathbb K}(\lambda) = {\cal L}(\lambda-\Theta)\  (K(\lambda)\otimes {\mathbb I})\  \hat {\cal L}(\lambda +\Theta), \label{gensol} \ee where $K$ is the c-number solution of the reflection equation, $\Theta$ some times is called inhomogeneity and henceforth for simplicity we shall consider it to be zero. In fact, in ${\mathbb K}$ one recognizes a `dynamical' (quantum impurity type) solution of the reflection equation (see e.g. \cite{doma}, \cite{zab}--\cite{dopa}). The entries of ${\mathbb K}$ are elements of the so called reflection algebra ${\mathbb R}$ with exchange relations dictated by the algebraic constraints (\ref{re}), (see also \cite{sklyanin}, \cite{mora}). It is clear that the general solution (\ref{gensol}) allows the  asymptotic expansion as $\lambda \to \infty$ providing the explict form of the reflection algebra generators, as we shall see in subsequent sections. The first order of such expansion (in the homogeneous gradation) yields the generators of the boundary quantum algebra ${\cal B}(U_{q}(gl_{n}))$, associated to $U_{q}(gl_{n})$,  which obey commutation relations dictated by the defining relations (\ref{re}) as $\lambda_{i} \to \infty$. The boundary quantum algebra is essentially a subalgebra of  $U_{q}(gl_{n})$ and ${\mathbb R}$, and provides the underlying algebraic structure in reflection equation exactly as quantum groups do in the Yang--Baxter equation. 

One may easily show that all the elements of the reflection algebra `commute' with the solutions of the reflection equation (see also \cite{dema}). In particular, by acting with the evaluation representation on the second space of (\ref{gensol}) it follows: \be  (\mbox{id} \otimes \pi_{\pm \lambda}){\mathbb K}(\lambda') = R(\lambda' \mp \lambda)\ (K(\lambda')\otimes {\mathbb I})\ \hat R(\lambda' \pm \lambda). \ee Now recalling the reflection equation (\ref{re}) and because of the form of the above expressions it is straightforward to show that \be (\mbox{id} \otimes \pi_{\lambda}){\mathbb K}(\lambda')\ ({\mathbb I} \otimes K(\lambda))= ({\mathbb I} \otimes K(\lambda))\  (\mbox{id} \otimes \pi_{-\lambda}){\mathbb K}(\lambda')\ \ee and consequently the entries of ${\mathbb K}$ in the evaluation representation `commute' with the c-number $K$ matrix (\ref{k}) \be \pi_{\lambda}({\mathbb K}_{ij}(\lambda'))\ K(\lambda) =  K(\lambda)\ \pi_{-\lambda}({\mathbb K}_{ij}(\lambda')), ~~~i,\j \in \{1, \ldots,n \}. \label{bcomm}\ee In fact, we conclude that any solution of the reflection equation commutes with the elements of the reflection algebra and the opposite. Relations analogous to (\ref{bcomm}) can be deduced for the $K$ matrix in the principal gradation (\ref{kp}) by multiplying (\ref{bcomm}) with ${\cal V}(\lambda)$ from the left and right. Equations (\ref{bcomm}) may be thought of as the boundary analogues of the `commutation' relations (\ref{inter}). 

The reflection algebra is also endowed with a coproduct inherited essentially from ${\cal A}$. In particular, let us first derive the coproduct of $\hat {\cal L}$, which is a consequence of (\ref{funda}) i.e. \be (\mbox{id} \otimes \Delta)  \hat {\cal L}(\lambda) =  \hat {\cal L}_{12}(\lambda)\ \hat {\cal L}_{13}(\lambda) \rightarrow \Delta(\hat {\cal L}_{ij}(\lambda)) = \sum_{k=1}^n \hat {\cal L}_{ik}(\lambda) \otimes \hat  {\cal L}_{kj}(\lambda)~~~~i,~j\in\{1, \ldots, n\}. \label{cob2} \ee It is then clear from (\ref{cob1}), (\ref{cob2}) that the elements of ${\mathbb R}$ form coproducts $\Delta: {\mathbb R} \to {\mathbb R} \otimes {\cal A}$, such that (see also \cite{dema}) \be \Delta({\mathbb K}_{ij}(\lambda)) = \sum_{k,l=1}^n{\mathbb K}_{kl}(\lambda) \otimes {\cal L}_{ik}(\lambda)\ \hat {\cal L}_{lj}(\lambda)~~~~i,~j \in \{1, \ldots, n\}. \label{coc} \ee Our main aim of course is to build  the corresponding quantum system that is the open quantum spin chain. The open spin chain may be constructed following the generalized QISM, introduced by Sklyanin \cite{sklyanin}. To achieve that we shall need tensor product realizations of the general solution (\ref{gensol}). We first need to define \be \hat T_{0}(\lambda) = (\mbox{id} \otimes \Delta^{(N)}) \hat {\cal L}(\lambda) =  \hat {\cal L}_{01}(\lambda)\ldots  \hat {\cal L}_{0N}(\lambda) \label{th} \ee then the  general tensor type solution of the (\ref{re}) takes the form \be {\cal T}_{0}(\lambda) = T_{0}(\lambda)\ K_{0}^{(r)}(\lambda)\ \hat T_{0}(\lambda),  \label{transfer0} \ee where the corresponding entries are simply coproducts of the elements of the reflection algebra, namely \be {\cal T}_{ij}(\lambda)= \Delta^{(N)}({\mathbb K}_{ij}(\lambda)). \ee Notice that relations similar to (\ref{bcomm}) may be derived for the more general solution of the reflection equation  
${\cal T}$\footnote{The operator ${\cal T}$ in principal and homogeneous gradation are related via the gauge transformation (\ref{gauge}), (\ref{vk}) \be {\cal T}_{0}^{(p)}(\lambda)={\cal V}_{0}(\lambda)\ {\cal T}^{(h)}_{0}(\lambda)\ {\cal V}_{0}(\lambda) \label{tp} \ee}. By virtue of these relations the commutation of the transfer matrix with certain non-local charges will be shown in a subsequent section. The derivation of the generalized intertwining relations follows essentially from the reflection equation (\ref{re}), the intertwining relations (\ref{bcomm}), and the form of the tensor solution ${\cal T}$.\\
\\
{\bf Proposition 3.3.} {\it Generalized linear intertwining relations for the ${\cal T}$ matrix (\ref{transfer0}) are valid, i.e. \be (\pi_{\lambda} \otimes
\mbox{id}^{\otimes N})\Delta^{'(N+1)}({\mathbb K}_{ij}(\lambda'))\ {\cal
T}(\lambda) = {\cal T}(\lambda)\ (\pi_{-\lambda} \otimes
\mbox{id}^{\otimes N})\Delta^{'(N+1)}({\mathbb K}_{ij}(\lambda')). \label{it0} \ee} {\it Proof:}
Intertwining relations for the monodromy matrix have been already established in (\ref{intert}), in fact it is clear due to the form of (\ref{funda}) that these relations hold for all elements ${\cal L}_{ij}(\lambda')$, $\hat {\cal L}_{ij}(\lambda')$. Analogous relations may be obtained for the $\hat T$
matrix. Indeed, recall that $\hat {\cal L}(\lambda) ={\cal L}^{-1}(-\lambda)$, then using (\ref{mono}) and (\ref{th}) we have \be \hat T(\lambda) = T^{-1}(-\lambda), \ee and  we conclude that \be (\pi_{-\lambda}
\otimes \mbox{id}^{\otimes N})\Delta^{(N+1)}(x)\ \hat T(\lambda) = \hat T(\lambda)\ (\pi_{-\lambda} \otimes \mbox{id}^{\otimes N})\Delta^{'(N+1)}(x), ~~x \in \{ {\cal L}_{ij}(\lambda'),\ \hat {\cal L}_{ij}(\lambda') \}. \label{im2} \ee  Having established the fundamental relations (\ref{intert}), (\ref{im2}) for the monodromy matrices $T$ and $\hat T$ we can now turn to the boundary case and establish similar relations for the operator ${\cal T}$.
   
From equations (\ref{intert}), (\ref{im2}) and because of the form of the elements of the reflection algebra (\ref{gensol}) it follows that \be
&&(\pi_{\lambda} \otimes \mbox{id}^{\otimes N})\Delta^{'(N+1)}({\mathbb K}_{ij}(\lambda'))\ T(\lambda) = T(\lambda)\ (\pi_{\lambda} \otimes
\mbox{id}^{\otimes N})\Delta^{(N+1)}({\mathbb K}_{ij}(\lambda')), \non\\
&&(\pi_{-\lambda} \otimes \mbox{id}^{\otimes N})\Delta^{(N+1)}({\mathbb K}_{ij}(\lambda'))\ \hat T(\lambda) = \hat T(\lambda)\ (\pi_{-\lambda} \otimes \mbox{id}^{\otimes N})\Delta^{'(N+1)}({\mathbb K}_{ij}(\lambda')).  \label{itt} \ee  Furthermore, due to (\ref{bcomm}), (\ref{coc}) it is clear that \be (\pi_{\lambda} \otimes \mbox{id}^{\otimes N})\Delta^{(N+1)}({\mathbb K}_{ij}(\lambda'))\ K(\lambda) = K(\lambda)\ (\pi_{-\lambda} \otimes \mbox{id}^{\otimes N})\Delta^{(N+1)}({\mathbb K}_{ij}(\lambda')). \label{iik} \ee Finally, using the equations (\ref{itt}), (\ref{iik}) combined with (\ref{transfer0}) one immediately obtains (\ref{it0}). Relations (\ref{it0}) hold also for ${\cal T}$ in the principal gradation (with $\pi_{\lambda} \to \tilde \pi_{\lambda}$). \finproof\\
\\
We can now introduce the transfer matrix of the open spin chain \cite{sklyanin}, which may be written as \be t(\lambda) = Tr_{0}\   \Big \{ M_{0}\ K_{0}^{(l)}(\lambda)\ {\cal T}_{0}(\lambda) \Big \}.  \label{transfer00} \ee $K^{(r)}$ is a solution of the reflection equation (\ref{re}) and $~K^{(l)}(\lambda) = K(-\lambda -i \mu{n\over 2})^{t}$ with $K$ being also a solution of (\ref{re}), not necessarily of the same type as $K^{(r)}$, and $M$ is defined in (\ref{M}). 

It can be proved using the fact that ${\cal T}$ is a solution of the reflection equation (\ref{re}) that \cite{sklyanin} \be \Big [ t(\lambda),\ t(\lambda') \Big ]=0 \ee  which ensures that the open spin chain derived by (\ref{transfer00}) is also integrable. Notice that as in the periodic case our construction is purely algebraic since the `quantum spaces' are not represented, but they are copies of ${\cal A}$. \\
\\
$\blacklozenge$ {\it The Hamiltonian:} It is useful to write down the Hamiltonian of the open spin chain. For this purpose we should restrict our attention in the case where the evaluation representation acts on the quantum spaces as well and $~{\cal L}(\lambda) \to R(\lambda)$, $~\hat {\cal L}(\lambda) \to \hat R (\lambda) = R^t(\lambda)~$ ($t$ denotes total transposition).  The Hamiltonian is proportional to ${d \over d \lambda} t(\lambda)\vert_{\lambda =0}$ \cite{done}, in particular we choose to normalize as: 
\be {\cal H} = -{(\sinh i\mu)^{-2N+1} \over {4 x(0)}} \Big( tr_{0} M_{0} \Big )^{-1}\ 
\ {d \over d \lambda} t(\lambda)\vert_{\lambda =0} \label{H0}. \ee  Here we consider $K^{(l)} ={\mathbb I}$, also $R$ is given by (\ref{sol2}), $K^{(r)}=K$ by (\ref{ansatz}) (homogeneous gradation) and 
$\rho$ is the representation derived in Proposition 3.2.\\ Then having in mind that  $~~R(0)= \hat R(0)=\sinh i \mu \ {\cal P}, ~~~K(0) = x(0)\ {\mathbb I}~~$ and also define \be H_{kl}= -{1\over 2 }{d\over d\lambda}({\cal P}_{kl}\ R_{kl}(\lambda)) \ee we may rewrite the Hamiltonian as
\be {\cal H} = \sum_{l=1}^{N-1} H_{l\ l+1}  
- {\sinh i\mu \over 4  x(0)}  \left( 
{d \over d \lambda} K_{ 1}(\lambda) \right) \Big \vert_{\lambda =0} + {tr_{0}\ M_{0}\ H_{N 0}  \over tr_{ 0}\ M_{ 0}}. \label{H} \ee Finally by taking into account that \be && H_{l\ l+1} = -{1 \over 2} (\cosh i \mu\ \rho(1) +\rho({\cal U}_{l}) ), ~~~{tr_{0}\ M_{0}\ H_{N0} \over tr_{0}M_{0}} =-{1\over 2} \cosh i\mu\  \rho(1) -{1\over 2} {tr_{0}\ M_{0}\ U_{N0}  \over tr_{0}M_{0}}, \non\\ &&\mbox{where} ~~~{tr_{0}\ M_{0}\ U_{N0} \over tr_{0}M_{0}} = c_{0}  \rho(1) \label{c} \ee ($c_{0}$ is a constant depending on $n$, and $U_{N0} =U_{0N}(q \to q^{-1})$ (\ref{sol}) see Appendix B) we conclude that the Hamiltonian (\ref{H}) can be written as
\be  {\cal H} = -\frac{1}{2}
\sum_{l=1}^{N-1} \rho({\cal U}_{l}) 
- \frac{\sinh i\mu\ y'(0)}{4 x(0)} \rho({\cal U}_{0})+ c \rho(1) \label{ht} \ee $c =  - \frac{\sinh i\mu\  x'(0)}{4 x(0)} -\frac{N}{2}\cosh i\mu  - {c_{0}\over 2}$. What is interesting in expression (\ref{ht}) is that it is written in terms of the generators of the ${\cal CB}_{N}$ in the representation of Proposition 3.2. This will be used later while studying the symmetry of the open spin chain. 
 
\subsection{Boundary non-local charges}

The presentation of section 3.3 relies mainly on abstract algebraic considerations, and as such it does not offer explicit expressions of the algebra generators, and consequently of conserved quantities that determine the symmetry of the open spin chain (\ref{transfer00}). To obtain explicit expressions of such quantities we need to investigate the asymptotic behaviour of  ${\cal T}$ (\ref{transfer0}) once the integrable boundary (\ref{k}) is implemented.
As in the bulk case the homogeneous and principal gradation will be examined and representations of generators of the reflection algebra will be obtained. Recall that the boundary non--local charges were derived for the $U_{q}(\widehat{sl_{2}})$ case in \cite{mene, dema, doikou}, and it turns out that they also generate the tridiagonal integrable structures as originally shown in \cite{pascal2}. We shall hereafter consider $K^{(r)}(\lambda) = K(\lambda)$ with $K$ given by (\ref{k}) and (\ref{kp}) for the homogeneous and principal gradation correspondingly.\\
\\
$\blacklozenge$ {\it Homogeneous gradation:} Recall that as $\lambda\to \infty$ the asymptotic behaviour of the monodromy matrix is given by (\ref{+1}). The asymptotics of $\hat {\cal L}(\lambda \to \infty)$ on the other hand is provided by \be \hat {\cal L}(\lambda \to \infty) \propto \hat {\cal L}^{+} =  \sum_{i>j} e_{ij}\otimes \hat t_{ij}  \label{th2} \ee where the elements $\hat t_{ij}$ and their corresponding coproducts are derived in Appendix A (\ref{ns2}). 
Then it follows that the asymptotics of $\hat T$ becomes (again here for simplicity the `auxiliary' space index $0$ is suppressed from $\hat T$ (\ref{th}) and ${\cal T}$ (\ref{transfer0}))
\be \hat T( \lambda\to \infty) \propto  \hat T^{+(N)} \ee with $\hat T^{+(N)}$ being a lower triangular matrix with entries belonging to ${\cal A}^{\otimes N}$
\be &&\hat  T_{ij}^{+(N)} =\Delta^{(N)}(\hat t_{ij}), ~~~i \geq j, ~~~~~\hat  T_{ij}^{+(N)}=0, ~~~i<j, ~~~~~i,j \in \{1, \ldots, n \}. \label{ee1} \ee 

The asymptotic behaviour of the $K$ matrix (\ref{k}) is also needed (also here we consider the arbitrary boundary parameter $m,\ \zeta$ to be finite) \be K( \lambda\to \infty) \propto K^{+}, \ee we write the non-zero entries of $K^{+}$
\be K_{11}^{+} = 2 \cosh i\mu m, ~~~K_{1n}^{+}=K_{n1}^{+}= -i,~~~K_{jj}^{+}=
e^{i\mu m}, ~~~j \in \{2,\ldots, n-1 \}. \label{kas0} \ee We come now to our main aim, which is the formulation of the asymptotics of the operator ${\cal T}$ (\ref{transfer0}), \be {\cal T}(\lambda\to \infty) \propto {\cal T}^{+(N)}. \ee Then from (\ref{transfer0}) and putting together (\ref{reppr}), (\ref{ee1}) and (\ref{kas0}) we conclude that the non-zero entries of ${\cal T}^{+(N)}$ are given by \be && {\cal T}_{11}^{+(N)} =2 \cosh i\mu m\ T_{11}^{+(N)}\ \hat T_{11}^{+(N)} -i T_{1 n}^{+(N)}\ \hat T_{11}^{+(N)} -i T_{11}^{+(N)}\ \hat  T_{n1}^{+(N)}+e^{i\mu m}  \sum_{j=2}^{n-1}  T_{1j}^{+(N)}\ \hat T_{j1}^{+(N)} \non\\ && {\cal T}_{1i}^{+(N)}=e^{i\mu m}  \sum_{j=i}^{n-1} T_{1j}^{+(N)}\ \hat T_{ji}^{+(N)} -i T_{11}^{+(N)}\ \hat T_{ni}^{+(N)},~~~  {\cal T}_{i1}^{+(N)}=e^{i\mu m}  \sum_{j=i}^{n-1} T_{ij}^{+(N)}\ \hat  T_{j1}^{+(N)} -i T_{in}^{+(N)}\ \hat  T_{11}^{+(N)}, \non\\ && i \in \{2, \ldots ,n \} \non\\
&& {\cal T}_{kl}^{+(N)}=e^{i\mu m}\sum_{j=max(k,l)}^{n-1} T_{kj}^{+(N)}\ \hat T_{jl}^{+(N)}, ~~~k,l \in \{ 2, \ldots, n-1\}. \label{trep2} \ee 
Notice that ${\cal T}_{ij}^{+(N)}$ are written exclusively in terms of tensor product realizations of the $U_{q}(gl_{n})$ elements, in particular ${\cal T}_{kl}^{+(N)} \in U_{q}(gl_{n-2}) ~~k,l \in \{ 2, \ldots, n-1\}$. The quantities presented in (\ref{trep2}), the boundary non--local charges, form the boundary quantum algebra ${\cal B}(U_{q}(gl_{n}))$. They are in fact its coproduct realizations. In particular, the corresponding exchange relations among the entries of ${\cal T}^{+}$ (\ref{trep2}) are entailed from (\ref{re}) as $\lambda_{i} \to \infty$, (no $\lambda$ dependence, homogeneous gradation) \begin{equation} R^{\pm}_{12}\ {\cal T}_{1}^{+(N)}\ \hat R_{12}^{+}\ {\cal T}_{2}^{+(N)} =  {\cal T}^{+(N)}_{2}\  R_{12}^{+}\ {\cal T}_{1}^{+(N)}\ \hat R_{12}^{\pm} \label{rr} 
\end{equation} where recall $\hat R^{\pm} ={\cal P} R^{\pm} {\cal P}$. In fact, the later formula (\ref{rr}) is an immediate consequence of the affine Hecke algebraic relation (\ref{braid2}). More specifically, to obtain (\ref{rr}) we consider $\rho(g_{1}) =U_{12}+q$, where $U$ is given in (\ref{sol}), we act on (\ref{braid2}) with ${\cal P}$ from the left and right, and set $~R_{12}^{\pm} = {\cal P}_{12}\ \rho(g_{1})^{\pm 1}$, also ${\cal T}_{1}^{+(N)}$ is a tensor representation of the generator $g_{0}$. As we shall see later in section 5 the charges (\ref{trep2}) commute with the open transfer matrix for special choice of the left boundary. This is the first time, to our knowledge, that explicit expressions of non-local charges associated to non-diagonal boundaries arising from (\ref{re}) are derived for the $U_{q}(\widehat{gl_{n}})$ case.\\
\\
$\blacklozenge$ {\it Principal gradation:} We come now to the study of the ${\cal T}$ asymptotics in the principal gradation. As in the bulk case to derive the boundary charge associated to the affine generators of ${\cal A}$, it is convenient to consider the principal gradation. It suffices to focus here on the simplest non trivial case i.e. the $U_{q}(\widehat{gl_{3}})$, in order to derive the `affine' boundary charge.

The asymptotic behaviour of $T$ (\ref{mono}) in the principal gradation  is given by (\ref{astt2}), (\ref{astt3}). The asymptotics of $\hat {\cal L}$ for the general $U_{q}(\widehat{gl_{n}})$ case is accordingly given by (keeping up to $e^{-{2  \lambda \over n}}$ terms) \be \hat {\cal L}(\lambda \to \infty ) \propto (D^{+} +e^{-{2  \lambda \over n}} \hat  B^{+}+\ldots) \ee with the non-zero entries of $\hat B^{+}$ given by
\be \hat B^{+}_{i+1\ i}=  \hat t_{i+1\ i}, ~~~~~\hat B^{+}_{1n}=\hat t^0_{1n}, \label{bbb}\ee
and consequently
\be \hat T(\lambda\to \infty) \propto  ( D^{+(N)} +e^{-{2  \lambda \over n}} \hat  B^{+(N)}+\ldots) \label{astt4} \ee the non-zero entries of $ \hat B^{+(N)}$ are also elements of ${\cal A}^{\otimes N}$  defined below \be \hat B^{+(N)}_{i+1\ i}=\hat T^{+(N)}_{i+1\ i}, ~~~~~\hat B^{+(N)}_{1n}=\Delta^{(N)}(\hat t^0_{1n}), \label{bb}\ee  $\hat T^{+(N)}_{ij}$ are given by (\ref{ee1}).  We shall also need the asymptotic behaviour of the $K$ matrix associated to $U_{q}(\widehat{gl_{3}})$ in the principal gradation (\ref{kp}) 
\be K^{(p)}(\lambda\to \infty) \propto (k+e^{-{2 \lambda \over 3}}f+\ldots) \label{kas} \ee with
\be  k= \left(
\begin{array}{ccc}
0        &0           &-i  \\
0        &e^{i \mu m} &0            \\
-i       &0           &0  \\
\end{array} \right)\,, ~~f= \left(
\begin{array}{ccc}
0        &0           &0  \\
0        &0           &0  \\
0        &0           &-2\cosh 2 i\mu \zeta  \\
\end{array} \right)\,.  \label{ask2} \ee Combining together equations (\ref{astt2}), (\ref{astt4}) for $n=3$, and (\ref{kas}), (\ref{ask2}) we conclude that  ${\cal T}( \lambda\to \infty)$ behaves as: \be {\cal T}( \lambda\to \infty) \propto ( D^{+(N)}\ k\  D^{+(N)} +e^{-{2\lambda \over 3}}( D^{+(N)}\ k\ \hat  B^{+(N)} +B^{+(N)}\ k\  D^{+(N)} + D^{+(N)}\ f\ D^{+(N)})+\ldots). \label{asypp} \ee Recalling the expressions of $ D^{+(N)}$, $ B^{+(N)}$, $\hat  B^{+(N)}$, $k$ and  $f$ we finally have
\be {\cal T}(\lambda\to \infty) \propto  ( \left(
\begin{array}{ccc}
0        &0           &{\cal T}_{13}^{+(N)}  \\
0        & {\cal T}_{22}^{+(N)}           &0  \\
{\cal T}_{31}^{+(N)}        &0           &0  \\
\end{array} \right) + e^{-{2 \lambda \over 3}} \left(
\begin{array}{ccc}
0        &{\cal T}_{12}^{+(N)}           &0  \\
{\cal T}_{21}^{+(N)}       &0           &0  \\
0        &0           &{\cal T}_{33}^{+(N)} \\
\end{array} \right) ) \ee ${\cal T}_{1j}^{+(N)}$, ${\cal T}_{j1}^{+(N)}$, ${\cal T}_{22}^{+(N)}$ are given by (\ref{trep2}) and for any $n$ we define \be {\cal T}_{nn}^{+(N)} = -2 \cosh 2 i \mu \zeta (T_{nn}^{+(N)})^{2} -i T_{nn}^{+(N)}\ \hat  B_{1n}^{+(N)}-i B^{+(N)}_{n1}\hat T_{nn}^{+(N)}. \label{taf} \ee  
Let us stress that the derivation of the boundary non-local charges is based on purely algebraic grounds, hence the form of the boundary non-local charges (\ref{trep2}), (\ref{taf}) is independent of the choice of representation. By keeping higher order terms in the expansion (\ref{asypp}) other boundary non-local charges will be entailed, however this case will not be examined here (see e.g. \cite{pascal2, doikou2}).

\section{The symmetry for special boundary conditions}

In the previous section we were able to derive the  boundary non-local charges as coproducts of the boundary quantum algebra (\ref{trep2}). We may simply write the abstract generators of the boundary quantum algebra ${\cal B}(U_{q}(gl_{n}))$, corresponding to the particular choice of $c$-number $K$ matrix (\ref{k}), using (\ref{trep2}), for $N=1$ i.e.\be && {\cal Q}_{11} =2 \cosh i\mu m\  t_{11}\ \hat t_{11} -i t_{1 n}\ \hat t_{11} -i t_{11}\ \hat  t_{n1}+e^{i\mu m}  \sum_{j=2}^{n-1} t_{1j}\ \hat t_{j1} \non\\ && {\cal Q}_{1i}=e^{i\mu m}  \sum_{j=i}^{n-1}t_{1j}\ \hat t_{ji} -i t_{11}\ \hat t_{ni},~~~  {\cal Q}_{i1}=e^{i\mu m}  \sum_{j=i}^{n-1} t_{ij}\ \hat t_{j1} -it_{in}\ \hat t_{11}, ~~~i \in \{2, \ldots , n\} \non\\
&& {\cal Q}_{kl}=e^{i\mu m}\sum_{j=max(k,l)}^{n-1} t_{kj}\ \hat t_{jl}, ~~~k,l \in \{ 2, \ldots, n-1\}. 
\label{genb} \ee 
notice in particular that ${\cal Q}_{kl} \in U_{q}(gl_{n-2}) \subset U_{q}(gl_{n}) ~~k,l \in \{ 2, \ldots, n-1\}$, with the generators of  $U_{q}(gl_{n-2})$ being  \be e_{i},~~f_{i},~~~~i \in \{2, \ldots, n-2 \} ~~~\mbox{and}~~~q^{\epsilon_{i}} ~~~~i\in \{2, \ldots, n-1\}. \label{n-2}\ee Furthermore, from (\ref{taf}) one more charge ${\cal Q}_{nn}$, associated with the affine generators of ${\cal A}$ may be obtained \be {\cal Q}_{nn} = -2 \cosh 2 i \mu \zeta\ t_{nn}\ \hat t_{nn} -it_{nn}\  \hat t^0_{1n}-i t^0_{n1}\ \hat t_{nn} \label{genb2}. \ee
For the following we shall denote $\hat {\cal B} = \{ {\cal Q}_{ij} \}\subseteq {\mathbb R}$. 
The elements ${\cal Q}_{ij}$ are equipped with a 
coproduct $\Delta: \hat {\cal B} \to \hat {\cal B} \otimes {\cal A}$ (see also (\ref{coc})). In particular, $\Delta({\cal Q}_{ij})$ are obtained from (\ref{genb}), (\ref{genb2})  with $t_{ij},~\hat t_{ij},~\to ~ \Delta(t_{ij}),~\Delta(\hat t_{ij})$. Bearing in mind expressions (\ref{cop2}), we may rewrite the $L$  coproducts in a more convenient for what follows form given in Appendix C. It is clear  via (\ref{genb}) and (\ref{reppr}), (\ref{ee1}), (\ref{trep2}), (\ref{taf}) that \be {\cal T}_{ij}^{+(N)} = \Delta^{(N)}({\cal Q}_{ij}). \label{repb} \ee From relations (\ref{genb}), by virtue of (\ref{eval}) and by defining \be \hat h_{i} = \pi_{0}(h_{i}), ~~~~~i\in \{1, \ldots, n \}\ee it can be deduced 
that \be \pi_{\lambda}({\cal Q}_{11}) &=& -2i\sinh i\mu\ q^{(\hat e_{11} +\hat e_{nn})}(-{\cosh i\mu m \over i \sinh i\mu} q^{-\hat h_{n}} + \hat e_{n1} +\hat e_{1n}  +2i \sinh i \mu\   e^{i\mu m} \sum_{j=2}^{n-1}\hat e_{jj}) \non\\ \pi_{\lambda}({\cal Q}_{1i})&=& -2i\sinh i\mu\ (ie^{i\mu m} \hat e_{i1} +\hat e_{in}) \non\\ \pi_{\lambda}({\cal Q}_{i1})&=& -2 i\sinh i\mu\ (ie^{i\mu m} \hat e_{1i} +\hat e_{ni}), ~~~i\in \{ 2, \dots, n-1 \} \non\\  \pi_{\lambda}({\cal Q}_{1n}) &=&\pi_{\lambda}({\cal Q}_{n1}) =-i q^{(\hat e_{11}+\hat e_{nn})}\non\\ \pi_{\lambda}({\cal Q}_{nn})   &=& -2i\sinh i\mu\ q^{(\hat e_{11} +\hat e_{nn})}({\cosh 2i\mu \zeta \over i \sinh i\mu} q^{\hat h_{n}} +e^{-2 \lambda}  \hat e_{n1} +e^{2 \lambda} \hat e_{1n})  \label{Q2} \ee notice also that $\pi_{\lambda}({\cal Q}_{1i}) = (\pi_{\lambda}({\cal Q}_{i1}))^t$. We shall also need the representations of the generators of $U_{q}(gl_{n-2})$ (\ref{n-2}), which are given by (\ref{eval}). The latter charges (\ref{Q2}) are similar, to the ones obtained in \cite{neg} for the open $U_{q}(\widehat{gl_{n}})$ case with non-diagonal boundaries. In particular, the charges of \cite{neg} may be simply expressed as linear combinations of (\ref{Q2}) and the evaluation representation of the $U_{q}(gl_{n-2})$ generators (\ref{n-2}). The parametrization used in \cite{neg} is associated explicitly with the one of \cite{abad}. Recall that relations among the parameters used in \cite{abad} (see also (\ref{ar})), and the parameters appearing in (\ref{kp}) are also provided in (\ref{rest2}). Note that in \cite{neg} the charges were essentially conjectured, whereas here they are explicitly derived as the evaluation representations of the abstract objects (\ref{genb}), (\ref{genb2}).

It is clear from the analysis of the previous section that ${\mathbb K}(\lambda' \to \infty)$ provide the charges (\ref{genb}), (\ref{genb2}). Therefore,  from (\ref{bcomm}) it is immediately entailed,
\be \pi_{\lambda}(y)\ K(\lambda)= K(\lambda)\
\pi_{-\lambda}(y),~~y\in \hat {\cal B}.
\label{ik} \ee  However, we also verified the validity of (\ref{ik}) by inspection. In fact, it can be explicitly shown that the $K$ matrix (\ref{k}) commutes with the $U_{q}(gl_{n-2})$ generators (\ref{n-2}) in the evaluation representation. Consequently $K$ commutes with the charges $\pi_{\lambda}({\cal Q}_{kl})$, $~k,\ l \in \{2, \ldots,n-2 \}$. Similarly, the intertwining relations with the rest of the charges (\ref{Q2}) may be verified by inspection. Since the $K$ matrix commutes with all the generators of $U_{q}(gl_{n-2})$ we may consider that $\hat {\cal B}$ consists essentially of $U_{q}(gl_{n-2})$ and the charges ${\cal Q}_{1i}$, ${\cal Q}_{1i}$ and ${\cal Q}_{nn}$. As an immediate consequence of (\ref{it0}) we may finally write, \be (\pi_{\lambda} \otimes
\mbox{id}^{\otimes N})\Delta^{'(N+1)}(y)\ {\cal
T}(\lambda) = {\cal T}(\lambda)\ (\pi_{-\lambda} \otimes
\mbox{id}^{\otimes N})\Delta^{'(N+1)}(y), ~~~y\in \hat {\cal B}. \label{it} \ee

The ultimate goal when dealing with a physical system such as a spin chain is to derive the existing symmetries, that is the set of conserved quantities. 
Having set all the necessary background we are now in the position to state the following propositions regarding the symmetry of the local Hamiltonian (\ref{ht}) and the algebraic transfer matrix (\ref{transfer0}), for special boundary conditions.\\
\\
{\bf Proposition 4.1.} {\it Representations of the boundary charges (\ref{trep2}) commute with the  generators of ${\cal CB}_{N}$ given in the representation of Proposition 3.2 i.e. \be  \Big [\rho({\cal U}_{l}),\ \pi_{0}^{\otimes N}({\cal T}^{+(N)}_{ij}) \Big ] =0, ~~~l \in \{ 0, \ldots ,N-1 \}. \label{comte2} \ee} {\it Proof:} Recall that all the boundary charges are expressed in terms of the $U_{q}(gl_{n})$ generators (see (\ref{q2}), (\ref{trep2})), thus via (\ref{comten}) it follows that (\ref{comte2}) is valid for all $l \in {1, \ldots N-1}$.\\
Now we need to prove (\ref{comte2}) for $l=0$.\\
$\bullet$ $N=1$: First it can be directly shown for $N=1$ by inspection via (\ref{eval}),  (\ref{solb}), (\ref{solb1}), and (\ref{n-2}) that $\rho({\cal U}_{0})$ is $U_{q}(gl_{n-2})$  invariant. Then immediately one deduces that (\ref{comte2}) is valid for all $~i,~j \in \{2, \ldots n-1 \}$. Recall that ${\cal T}^{+(N)}_{ij}$ $~i,~j \in \{2, \ldots n-1 \}$ (\ref{trep2}) are expressed exclusively in terms of the generators of $U_{q}(gl_{n-2})$ (\ref{n-2}).\\ It can be also shown by inspection for $N=1$ via (\ref{solb}), (\ref{solb1}) (Proposition 3.2) and (\ref{repb}), (\ref{Q2}) that (\ref{comte2}) holds for the rest of the boundary charges. \\
$\bullet$ $\forall\ N$: By means of (\ref{solb}), (\ref{solb1}), (\ref{repb}) and (\ref{bc2}) ($L \to N$) it follows that (\ref{comte2}) is valid for $l=0$ $~\forall\ N$. This is consequence of the fact that the elements in the first site of (\ref{bc2}), are elements of $\hat {\cal B}$, the representations of which on $\mbox{End}({\mathbb C}^{n})$ commute anyway with (\ref{solb}) (see (\ref{inter}). \finproof \\
\\
{\bf Corollary:} {\it The open spin chain Hamiltonian (\ref{ht}) commutes with  the representations of  the boundary charges (\ref{trep2}) \be \Big [{\cal H},\ \pi_{0}^{\otimes N}({\cal T}^{+(N)}_{ij}) \Big
] =0. \label{comh} \ee} This is evident from the form of ${\cal H}$ 
(\ref{ht}), which is expressed solely in terms of the representation $\rho({\cal U}_{l})$ of ${\cal CB}_{N}$.
 
The aim now is to study the symmetry displayed by the open algebraic transfer matrix (\ref{transfer0}) for particular choices of boundary conditions. An effective way to achieve that is to exploit the existence of the relations (\ref{it}). Indeed, equations (\ref{it}) turn out to play a crucial role, bearing explicit algebraic relations among the entries of the ${\cal T}$ matrix and the non--local charges (\ref{trep2}) (see also \cite{doikou}). We shall focus henceforth on the homogeneous gradation, however analogous results may be easily deduced for the principal gradation as well. Note also that in what follows we consider $K^{(l)} = {\mathbb I}$ (homogeneous gradation). In the case we choose a different left boundary we shall explicitly state it. \\ 
Let \be {\cal T}(\lambda)= \left(\begin{array}{ccccccc}
{\cal A}_{1} &{\cal B}_{12} &\cdots  &\cdots &\cdots &\cdots &{\cal B}_{1n}\\
{\cal C}_{21} &{\cal A}_{2} &{\cal B}_{23} &\ddots &\ddots &\ddots &{\cal B}_{2n} \\
\vdots & \ddots &\ddots &\ddots   &\ddots  & \ddots  &\vdots\\ {\cal C}_{i1} &\ldots &{\cal C}_{i\ i-1} &{\cal A}_{i} &{\cal B}_{i\ i+1} &\ldots &{\cal B}_{in}\\
\vdots & \ddots &\ddots  & \ddots  & \ddots  & \ddots  &\vdots\\ \vdots & \ddots & \ddots  & \ddots & \ddots & {\cal A}_{n-1} &{\cal B}_{n-1\ n}\\
{\cal C}_{n1} &\ldots &\cdots  &\cdots  &\cdots  &{\cal C}_{n\ n-1} &{\cal A}_{n}
\end{array}\right) ~~~\mbox{and} ~~~t(\lambda) =\sum_{j=1}^{n} q^{(n-2j+1)}{\cal A}_{j} \label{tr2} \ee Then via (\ref{it}) and (\ref{form1})--(\ref{form2b}) we can prove that:\\
\\
{\bf Proposition 4.2.} {\it The open transfer matrix (\ref{transfer0}), (\ref{tr2}) built with the $K^{(r)}$ matrix  (\ref{k})  is $U_{q}(gl_{n-2})$ invariant i.e., \be \Big [t(\lambda),\ U_{q}(gl_{n-2}) \Big ]=0. \ee}
{\it Proof:} Let us first introduce some useful notation, namely 
\be &&E_{ii}^{(N)} = \Delta^{(N)}(q^{\epsilon_{i}}), ~~~~H_{i}^{(N)}= \Delta^{(N)}(q^{h_{i}}), \non\\ && E_{i\ i+1}^{(N)} = \Delta^{(N)}(e_{i}), ~~~~E_{i+1\ i}^{(N)} = \Delta^{(N)}(f_{i}). \label{nota} \ee
Consider now (\ref{it}) for $y=e_{j},\ j=\{2, \ldots n-2 \}$ then by means of (\ref{form1}) we obtain the commutation relations:
\be &&\Big [ E_{j\ j+1}^{(N)},\ {\cal A}_{j}\Big ]= -q^{-{1\over 2}}(H_{j}^{(N)})^{-{1 \over 2}}{\cal C}_{j+1\ j}
~~\Big [E_{j\ j+1}^{(N)},\ {\cal A}_{j+1}\Big ]=  q^{{1\over 2}}{\cal C}_{j+1\ j}(H_{j}^{(N)})^{-{1 \over 2}}
 \non\\ && \Big [E_{j\ j+1}^{(N)},\ {\cal A}_{i} \Big ]= 0,~~~i \neq j, ~j+1.
\label{com2} \ee
Similarly,  for $y=f_{j},\ j=\{2, \ldots n-2 \}$ in (\ref{it}) and by means of
(\ref{form2}): \be &&\Big [ E_{j+1\ j}^{(N)},\ {\cal A}_{j}\Big ]= q^{-{1\over
2}}{\cal B}_{j\ j+1}(H_{j}^{(N)})^{-{1 \over 2}}, ~~\Big [E_{j+1\ j}^{(N)},\ {\cal
A}_{j+1}\Big ]= - q^{{1\over 2}}(H_{j}^{(N)})^{-{1 \over 2}}{\cal B}_{j\ j+1} \non\\ && \Big [E_{j+1\ j}^{(N)},\ {\cal A}_{i} \Big ]= 0,~~~i \neq j,  ~j+1. \label{com3} \ee Finally, for  $y=q^{\epsilon_{j}},\ j=\{2, \ldots n-1 \}$, and also for  $y=q^{\pm {h_{j} \over 2}},\ j=\{2, \ldots n-2 \}$ in (\ref{it})
we obtain via (\ref{form2b})  \be &&\Big [E_{jj}^{(N)},\ {\cal A}_{i}  \Big ]=0, ~~~i,\ j \in \{1,\ldots, n\}, ~~~ j \neq 1,~n \non\\ && q^{\pm{1\over 2}}(H_{j}^{(N)})^{\pm {1 \over 2}} {\cal B}_{j\ j+1} =  q^{\mp{1\over 2}} {\cal B}_{j\ j+1} (H_{j}^{(N)})^{\pm {1 \over 2}}, ~~~q^{\mp{1\over 2}}(H_{j}^{(N)})^{\pm {1 \over 2}} {\cal C}_{j+1\ j} =  q^{\pm{1\over 2}} {\cal C}_{j+1\ j} (H_{j}^{(N)})^{\pm {1 \over 2}} \non\\ && j \in \{2,\ldots, n-2\}. \label{com4} \ee Then from (\ref{com2}), (\ref{com3}) it follows that \be && \Big [E_{j\ j+1}^{(N)},\ \sum_{i=1}^{n}q^{(n-2i+1)}{\cal A}_{i} \Big ] = q^{(n-2j)}(-q^{{1\over 2}}(H_{j}^{(N)})^{-{1 \over 2}} {\cal C}_{j+1\ j} +q^{-{1\over 2}}{\cal C}_{j+1\ j}(H_{j}^{(N)})^{-{1 \over 2}}) \non\\ &&\Big [E_{j+1\ j}^{(N)},\ \sum_{i=1}^{n}q^{(n-2i+1)}{\cal A}_{i} \Big ] = q^{(n-2j)}(q^{{1\over 2}}{\cal B}_{j\ j+1} (H_{j}^{(N)})^{-{1 \over 2}}  -q^{-{1\over 2}}(H_{j}^{(N)})^{-{1 \over 2}}{\cal B}_{j\ j+1}) \label{com4b} \ee and by virtue of (\ref{tr2}), (\ref{com4}), (\ref{com4b}) we conclude that \be && \Big [t(\lambda),\ E_{
 j\ j+1}^{(N)}\Big]=\Big [t(\lambda),\ E_{j+1\ j}^{(N)}\Big]=0, ~~~j\in \{2, \ldots, n-2 \} \non\\ && \Big [t(\lambda),\ E_{j j}^{(N)}\Big]=0, ~~~j\in \{2, \ldots, n-1 \}. \label{cr1} \ee
Equations (\ref{cr1}) show that the transfer matrix commutes with the  coproduct representations of the $U_{q}(gl_{n-2})$ generators, and this concludes our proof. \finproof\\
In the case where $K^{(r)}={\mathbb I}$ the intertwining relations (\ref{it}) hold for all the generators of $U_{q}(gl_n)$, and therefore in the spirit of Proposition 4.2. it is straightforward to show that the transfer matrix is then $U_{q}(gl_{n})$ invariant, recovering the result of \cite{menes}.\\
\\
{\bf Proposition 4.3.} {\it The open transfer matrix (\ref{transfer0}) commutes with all the non-local charges ${\cal T}^{+(N)}_{ij}$ (\ref{trep2}), i.e. \be  \Big [t(\lambda),\ {\cal T}^{+(N)}_{ij} \Big ] =0. \label{symm} \ee}
{\it Proof:}  The proof relies primarily on the existence of the generalized intertwining relations (\ref{it}). In particular, the commutation of the transfer matrix with the last set ${\cal T}_{kl}^{+(N)}$, $k,\ l \in \{ 2, \ldots,  n-1\}$ of (\ref{trep2}) is a corollary of Proposition 4.2. Recall that these charges are expressed in terms of the generators of $U_{q}(gl_{n-2})$. \\
Now our aim is to show the commutation of $t(\lambda)$ with the rest of the charges. To illustrate the method we use the simplest example i.e. the $U_{q}(\widehat{gl_{3}})$. In appendix D explicit expressions for representations of the fundamental objects $\Delta^{'(N+1)}({\cal Q}_{ij})$ are provided. Then via (\ref{it}) for $y= {\cal Q}_{12}$, (\ref{tr2}) (for $n=3$)  and (\ref{form3}) we conclude \be && \Big [{\cal A}_{1},\ {\cal T}^{+(N)}_{12} \Big ] = -e^{i\mu m}w q^{-1} {\cal B}_{12} (E^{(N)}_{22})^2 , ~~~\Big [{\cal A}_{3},\ {\cal T}^{+(N)}_{12} \Big ] = i w {\cal C}_{32}\ E_{11}^{(N)}  E_{33}^{(N)}, \non\\ && \Big [{\cal A}_{2},\ {\cal T}^{+(N)}_{12} \Big ] = e^{i\mu m}\ w\ q^{-1}(E^{(N)}_{22})^2 {\cal B}_{12} -i q^{-1} w E_{11}^{(N)}  E_{33}^{(N)} {\cal C}_{32}, \non\\ && \Big [ {\cal T}^{+(N)}_{12},\ {\cal C}_{21} \Big] =i w q^{-1} E_{11}^{(N)} E_{33}^{(N)} {\cal C}_{31} -q^{-1}e^{i\mu m}w (E_{22}^{(N)})^{2}{\cal A}_{1} +q^{-1}e^{i\mu m} w {\cal A}_{2} (E_{22}^{(N)})^{2}. \label{com5} \ee 
Similarly,  via (\ref{it}) for $y={\cal Q}_{21}$ and (\ref{tr2}), (\ref{form3}) the following commutation relations are obtained:
\be && \Big [{\cal A}_{1},\ {\cal T}^{+(N)}_{21} \Big ] = e^{i\mu m} w q^{-1}
 (E^{(N)}_{22})^2 {\cal C}_{21} , ~~~\Big [{\cal A}_{3},\ {\cal T}^{+(N)}_{21} \Big] = -i w E_{11}^{(N)} E_{33}^{(N)} {\cal B}_{23}, \non\\ && \Big [{\cal A}_{2},\ {\cal T}^{+(N)}_{21} \Big ] = -e^{i\mu m} w q^{-1}{\cal C}_{21}\ (E^{(N)}_{22})^2 +i q^{-1} w {\cal B}_{23} E_{11}^{(N)} E_{33}^{(N)}, \non\\ && \Big [ {\cal T}^{+(N)}_{21},\ {\cal B}_{12} \Big]  =- i w q^{-1}{\cal B}_{13} E_{11}^{(N)} E_{33}^{(N)}  +q^{-1}e^{i\mu m}w {\cal A}_{1} (E_{22}^{(N)})^{2} + q^{-1}e^{i\mu m} w (E_{22}^{(N)})^{2} {\cal A}_{2}. \label{com6} \ee 
For $y={\cal Q}_{11}$ we have by means of (\ref{it}), (\ref{tr2}), (\ref{form3}):
\be && \Big [{\cal A}_{1},\ {\cal T}^{+(N)}_{11} \Big ] =-w q^{-1}{\cal B}_{12} {\cal T}^{+(N)}_{21} +i w q^{-1}{\cal B}_{13}\ E_{11}^{(N)} E_{33}^{(N)} +w q^{-1}{\cal T}^{+(N)}_{12}{\cal C}_{21} -i w q^{-1} E_{11}^{(N)} E_{33}^{(N)} {\cal C}_{31}, \non\\ && \Big [{\cal A}_{2},\ {\cal T}^{+(N)}_{11} \Big ] = -w q {\cal C}_{21} {\cal T}^{+(N)}_{12} +w q {\cal T}^{+(N)}_{21} {\cal B}_{12}+e^{i\mu m}w^{2} (E_{22}^{(N)})^{2} {\cal A}_{2} -e^{i\mu m}w^{2} {\cal A}_{2} (E_{22}^{(N)})^{2} \non\\ && \Big [{\cal A}_{3},\ {\cal T}^{+(N)}_{11} \Big ] =- i w q  E_{11}^{(N)} E_{33}^{(N)} {\cal B}_{13} +iw q {\cal C}_{31} E_{11}^{(N)} E_{33}^{(N)}.  \label{com7} \ee 
The following useful commutation relations are also valid, arising from (\ref{it}) for $y=t_{22}^2$ and $y =t_{11}\ t_{33}$,
\be &&  q\ E_{11}^{(N)}\ E_{33}^{(N)}\ {\cal C}_{32} ={\cal C}_{32}\ E_{11}^{(N)}\ E_{33}^{(N)} ,~~~(E_{22}^{(N)})^{2}\ {\cal B}_{12} = q^{2} {\cal B}_{12}\ (E_{22}^{(N)})^{2}, \non\\ &&  E_{11}^{(N)}\ E_{33}^{(N)}\ {\cal B}_{23} =q\ {\cal B}_{23}\ E_{11}^{(N)}\ E_{33}^{(N)} ,~~~ q^{2}(E_{22}^{(N)})^{2}\ {\cal C}_{21} = {\cal C}_{21}\ (E_{22}^{(N)})^{2} \non\\ && \Big [E_{11}^{(N)}\ E_{33}^{(N)},\ {\cal B}_{13}\Big ]= \Big [E_{11}^{(N)}\ E_{33}^{(N)},\ {\cal C}_{31} \Big ]= 0. \label{com9} \ee 
Combining equations (\ref{tr2}) and (\ref{cr1}), (\ref{com5})--(\ref{com9}) we conclude \be  \Big [t(\lambda),\ {\cal T}^{+(N)}_{12} \Big ] = \Big[t(\lambda),\ {\cal T}^{+(N)}_{21} \Big ] =\Big [t(\lambda),\ {\cal T}^{+(N)}_{11} \Big ] = 0. \label{com10} \ee The algebraic relations (\ref{com2})--(\ref{com4b}), (\ref{com5})--(\ref{com9}) and also (\ref{cr1}), (\ref{com10}) are of the most important results of this article. Although the general $U_{q}(\widehat{gl_{n}})$ case is technically more complicated, it may be treated  along the same lines. In particular, one may show recursively starting from ${\cal T}^{+(N)}_{1\ n-1}$ up to ${\cal T}^{+(N)}_{12}$ and finally for ${\cal T}^{+(N)}_{11}$ by exploiting relations of the type (\ref{cr1}), (\ref{com5})--(\ref{com9}) that 
\be \Big [t(\lambda),\ {\cal T}^{+(N)}_{1i} \Big ] =\Big [t(\lambda),\ {\cal T}^{+(N)}_{i1} \Big ] =\Big [t(\lambda),\ {\cal T}^{+(N)}_{11} \Big ] = 0. \label{gener} \ee  \finproof

We can also get commutation relations between the affine generator ${\cal T}^{+(N)}_{nn}$ (for any $n$) and the transfer matrix. Indeed it follows from (\ref{it}), (\ref{form4})   
\be && \Big [{\cal A}_{1},\ {\cal T}^{+(N)}_{nn} \Big ] =i\ w\ q\ e^{2 \lambda}{\cal B}_{1n}\  E_{11}^{(N)}\ E_{nn}^{(N)} -i\ w\ q\ e^{2 \lambda}\  E_{11}^{(N)}\ E_{nn}^{(N)}\ {\cal C}_{n1},  ~~~\Big [{\cal A}_{j},\ {\cal T}^{+(N)}_{nn} \Big ] = 0, \non\\ && j\in \{2, \ldots, n-1 \} \non\\ &&\Big [{\cal A}_{n},\ {\cal T}^{+(N)}_{nn} \Big ] =  i\ w\ q^{-1}\ e^{-2  \lambda} {\cal C}_{n1}\ E_{11}^{(N)}\ E_{nn}^{(N)}  -i\ w\ q^{-1}\ e^{-2 \lambda}  E_{11}^{(N)}\ E_{nn}^{(N)}\ {\cal B}_{1n},  \label{com8} \ee then by means of the following commutations \be \Big [ E_{11}^{(N)}\ E_{nn}^{(N)},\ {\cal B}_{1n} \Big ] = \Big [ E_{11}^{(N)}\ E_{nn}^{(N)},\ {\cal C}_{n1}\Big ]= 0 \label{com11} \ee and via (\ref{tr2}), (\ref{com8}), (\ref{com11}) we conclude \be \Big [t(\lambda),\ {\cal T}^{+(N)}_{nn} \Big ] = 2\ i\  w\ \sinh (2\lambda +in \mu)\  ({\cal B}_{1n}-{\cal C}_{n1})\ E_{11}^{(N)}\ E_{nn}^{(N)}. \label{com12} \ee The transfer matrix commutes only with the `non--affine' boundary generators, while it does not commute with the affine generator ${\cal T}^{+(N)}_{nn}$. This is also true when trivial boundary conditions are implemented i.e. $K^{(r)}={\mathbb I}$, in this case as well the affine generators of $U_{q}(\widehat{gl_{n}})$ do not commute with the transfer matrix \cite{menes}. In both cases the open spin chain enjoys the symmetry associated to the non--affine generators. Relations of the type (\ref{it}) may be also used for studying the symmetries discussed in \cite{done} for the special case where $K^{(r)}$ is diagonal. Let us briefly review what is known up to date on the symmetries of open spin chains, with $K^{(l)}(\lambda) ={\mathbb I}$ (homogeneous gradation):
\begin{itemize}
\item $K^{(r)}(\lambda) ={\mathbb I}$. The transfer matrix then enjoys the $U_{q}(gl_{n})$ symmetry as shown on \cite{kusk}, \cite{menes}. In fact this case may be easily studied in our framework as a limit of the solution (\ref{k}) as $|\zeta| \to \infty$.
\item $K^{(r)}(\lambda) =\mbox{diag}\Big (\underbrace{\alpha, \ldots , \alpha}_{\mbox{$l$}},\  \underbrace{\beta, \ldots , \beta}_{\mbox{$n-l$}} \Big )$\\ with
$\alpha(\lambda;\ \xi ) = \sinh (-\lambda +i\mu \xi)\ e^{\lambda}, ~~\beta(\lambda;\ \xi ) = \sinh (\lambda +i\mu \xi)\ e^{-\lambda}$, where $\xi$ is an arbitrary boundary parameter \cite{DVGR}. In this case it was shown in \cite{done} that the symmetry of the open transfer matrix is $U_{q}(gl_{l}) \otimes U_{q}(gl_{n-l})$. 
\item
$K^{(r)}(\lambda)$ is given by (\ref{k}), then as we proved in Propositions 4.2 and 4.3 the symmetry displayed by the transfer matrix is associated to the boundary non-local charges (\ref{trep2}).
\end{itemize}

Let us at this point deal with a different left boundary i.e. $K^{(l)}(\lambda) = K(-\lambda -i \mu{n\over 2}; \ i\mu m \to \infty)$ (\ref{k}), then we can explicitly write \be K^{(l)}(\lambda) \propto diag(e^{-2\lambda - i \mu n},\ e^{-2\lambda - i \mu n}, \ldots, e^{-2\lambda - i \mu n},\ e^{2\lambda + i \mu n} ). \label{new}\ee Recalling (\ref{tr2}) and also using (\ref{transfer0}) and (\ref{new}) we can write
\be t(\lambda) = e^{-2\lambda -i\mu} {\cal A}_{1} +  e^{2\lambda+ i\mu} {\cal A}_{n} + e^{-2\lambda}\sum_{j=2}^{n-1}q^{-2j+1}{\cal A}_{j}. \label{tt3} \ee Then by virtue of 
(\ref{com2})--(\ref{com4}), (\ref{com8}), (\ref{com11}) and (\ref{tt3}) we conclude that: \be  \Big [t(\lambda),\ {\cal T}_{nn}^{+(N)} \Big ]=0, ~~~~\Big [t(\lambda),\ U_{q}(gl_{n-2}) \Big ]=0. \label{fin} \ee Appropriate choice of the left boundary leads to the commutation of the transfer matrix with the non-local charge associated to the affine generators of $U_{q}(\widehat{gl_{n}})$. Although the $U_{q}(gl_{n-2})$ symmetry is still preserved in the presence of the non-trivial left boundary (\ref{new}), the transfer matrix does not commute anymore with each one of the charges ${\cal T}_{1j}^{+(N)}$, ${\cal T}_{j1}^{+(N)}$ (see e.g. (\ref{com5})--(\ref{com7})).  Let us point out that the discovered symmetries (\ref{cr1}), (\ref{gener}), (\ref{fin}) are independent of the choice of representation, and thus they have a universal character.
 
Similar commutation relations may be entailed for the transfer matrix in the principal gradation by virtue of the gauge transformation (\ref{v}). Recall the transfer matrix in the principal gradation is  \be t^{(p)}(\lambda) = Tr_{0}\  \Big \{ K_{0}^{(l)}(\lambda)\ {\cal
T}_{0}^{(p)}(\lambda) \Big \} \label{pp} \ee where ${\cal T}^{(p)}$ is given by
(\ref{tp}). In the case where $ K^{(l)}$ is diagonal the transfer matrix can be written as combination of the diagonal entries of ${\cal T}$.  However, the exchange relations between ${\cal A}_{j}$ and the charges ${\cal T}^{+(N)}_{ij}$ are given by (\ref{com2})--(\ref{com4}), (\ref{com5})--(\ref{com7}), then the  commutators  between $t^{(p)}$ and ${\cal T}^{+(N)}_{ij}$ may be immediately deduced. 

The implementation of a more general left boundary, $K^{(l)}$ non-diagonal (\ref{k}), leads in principle to a modified set of conserved quantities, and hence the symmetry of the system is modified. The treatment of the general case is straightforward along the lines described in this section. In particular, given a more general choice of non-diagonal left boundary, in both gradations, the transfer matrix may be written as a combination of all entries of ${\cal T}$ (\ref{tr2}). In this case one has to exploit the exchange relations between all the entries of ${\cal T}$ and ${\cal T}_{ij}^{+(N)}$ (see e.g. (\ref{com2})--(\ref{com4}), (\ref{com5})--(\ref{com9})) in order to derive the commutators between $t(\lambda)$ and the boundary non--local charges.

It is worth pointing out the significance of the method demonstrated here for the investigation of the open transfer matrix symmetry. Usually the study of the symmetry of an open transfer matrix (see e.g. \cite{done, menes}) relies primarily on the fact that the monodromy matrix $T$ ($\hat T$) reduces to upper (lower) triangular matrix as $\lambda \to \infty$, which facilitates enormously the algebraic manipulations. There exist however cases where the monodromy matrix does not reduce to such a convenient form as $\lambda \to \infty$ (see e.g. \cite{domapr}). In these cases the most effective way to investigate the corresponding symmetry is the method presented in \cite{dema}, and in the present article in the context of integrable open quantum spin chains. More specifically, one should derive in some way (e.g. by direct computation) linear intertwining relations of the type (\ref{it}), by means of which exchange relations between the entries of transfer matrix and the corresponding non--local charges can be deduced (see e.g. (\ref{com2})--(\ref{com4b}), (\ref{com5})--(\ref{com9}), (\ref{com8})--(\ref{com11})), enabling the derivation of the set of conserved quantities (see e.g. (\ref{symm}), (\ref{fin})). All the mentioned relations (\ref{com2})--(\ref{com4b}), (\ref{com5})--(\ref{com9}), (\ref{com8})--(\ref{com11}) are manifestly exchange relations of the reflection algebra, since (\ref{it0}), (\ref{it}) are immediate consequences of the defining equation (\ref{re}).

\section{Comments}

We would like to comment on some open problems that are worth pursuing in the spirit of the present analysis. 

The existence of other representations of quotients of the affine Hecke algebra that provide solutions to the reflection equation is also an interesting direction to be explored. For instance the generalization of the `twin' representation proposed in \cite{mawo, doma} is an intriguing problem. What makes this representation especially appealing is the fact that offers a novel approach on the understanding of boundary phenomena in the context of integrable systems as pointed out in \cite{doma, domapr}. In addition, such a  representation can not be attained in a straightforward manner by means of the quantum group scheme (for more details see e.g. \cite{doma, domapr}), an evidence that further supports the Hecke algebraic approach in solving the reflection equation. It would be also instructive to identify the representations of the affine Hecke algebra, that give rise to more general solutions of the reflection equation, for the $U_{q}(\widehat{gl_{n}})$ case, recently derived in \cite{lima, chin2}. 

Recall that here essentially we considered the left boundary trivial. Based on the affine Hecke algebra it would be interesting to generalize the analysis presented in \cite{nic} on the spectrum of the non-local charges (\ref{trep2}), for various values of the parameters of the $B$-type Hecke algebra. In the case where the left boundary of the spin chain is also non--trivial one needs to introduce an additional generator to the affine Hecke algebra, corresponding to the left boundary, and proceed generalizing the approach described in \cite{degi, degi2} to higher rank algebras.

A natural question raised is what is the analogue of the affine Hecke algebra in the case of SNP boundary conditions. A possible guess is that it should be a generalization of a Birman--Wenzl--Murakami type algebra \cite{bmw} including generators that correspond to the boundaries.
Note also that for both SP and SNP boundary conditions the derivation of higher non-local conserved charges is an intricate problem that needs to be further explored (see e.g. \cite{pascal2, doikou2}). In particular, the crucial point is to identify the higher rank generalization of the infinite dimensional algebra discovered in \cite{pascal3}, with elements (\ref{trep2}), (\ref{taf}) together with higher order non-local charges. 

Finally, the formulation of a field theory with SP boundary conditions is a problem worth investigating. As mentioned already in the introduction the known boundary conditions from the field theory point of view are the SNP ones. They were introduced and studied at the classical level in \cite{cor}, and the corresponding $K$ matrices were obtained for the first time in \cite{gand}. Also in \cite{dema} the corresponding boundary non-local charges and  $K$ matrices were derived. It is our intention to examine this case and derive directly the {\it quantum} non-local charges for the SNP boundary conditions from the spin chain point of view \cite{doikous}--\cite{doikoupr}, \cite{doikou2}. \\
We hope to address the aforementioned issues in full detail in forthcoming works.\\
\\
\textbf{Acknowledgements:} I am grateful to P.P. Martin for valuable discussions on (affine) Hecke algebras. I am also thankful to D. Arnaudon, J. Avan, L. Frappat and  E. Ragoucy for illuminating comments, and for bringing to my attention references \cite{difre, moras}. I would like finally to thank the organizers of the 6th Bologna Workshop on `CFT and Integrable models', where part of this work was presented. This work is supported by the TMR Network `EUCLID. Integrable models and applications: from strings to condensed matter', contract number HPRN-CT-2002-00325, and CNRS.

\appendix

\section{Appendix} 

In this appendix we shall introduce some useful quantities (elements of ${\cal A}$), and we shall also derive the corresponding coproducts.

It will be convenient for our purposes here to rewrite ${\cal E}_{ij}$ (\ref{ee}), in a slightly different form, and also introduce $\hat {\cal E}_{ij}$ defined by the recursive relations:\\ $\hat {\cal E}_{i\ i+1}=  e_{i}$ $~i\in \{1, \ldots, n-1 \}$, and for $|i-j|>1$
\be &&  {\cal E}_{ij} = {1\over (|i-j|-1)} \sum_{k=min(i,\ j)+1}^{max(i,\ j)-1}( {\cal E}_{ik}\  {\cal E}_{kj} -q^{\mp 1} {\cal E}_{kj}\  {\cal E}_{ik}), ~~~j \lessgtr k \lessgtr i \non\\ &&\hat {\cal E}_{ij} = {1\over(|i-j|-1)} \sum_{k=min(i,\ j)+1}^{max(i,\ j)-1}(\hat {\cal E}_{ik}\ \hat {\cal E}_{kj}
-q^{\pm 1}\hat {\cal E}_{kj}\ \hat {\cal E}_{ik}), ~~~j\lessgtr k \lessgtr i, \non\\ && i,~j \in \{ 1, \ldots , n\}. \label{q1} \ee 
Also define \be &&t_{ij} = 2\sinh i\mu\  q^{-{1\over 2}} q^{{\epsilon_{i} \over 2}}q^{{\epsilon_{j} \over 2}}\ {\cal E}_{ji}, ~~i<j, ~~~ t^{-}_{ij} = -2\sinh i\mu\  q^{{1\over 2}} q^{-{\epsilon_{i} \over 2}}q^{-{\epsilon_{j} \over 2}}\ {\cal E}_{ji}, ~~i>j \non\\ & &\hat t_{ij}= 2\sinh i\mu\  q^{-{1\over 2}} q^{{\epsilon_{i} \over 2}}q^{{\epsilon_{j} \over 2}}\ \hat {\cal E}_{ji}, ~~i>j, ~~~ \hat t^{-}_{ij} = -2\sinh i\mu\  q^{{1\over 2}} q^{-{\epsilon_{i} \over 2}}q^{-{\epsilon_{j} \over 2}}\ \hat{\cal E}_{ji}, ~~i<j , \non\\ && \hat t^0_{1n} = 2 \sinh i\mu\  q^{-{1\over 2}} q^{{\epsilon_{1} \over 2}}q^{{\epsilon_{n} \over 2}}\ e_{n} , ~~~t^0_{n1} = 2 \sinh i\mu\  q^{-{1\over 2}} q^{{\epsilon_{1} \over 2}}q^{{\epsilon_{n} \over 2}}\ f_{n}, \non\\ && t^{0-}_{1n} = -2 \sinh i\mu\  q^{{1\over 2}} q^{-{\epsilon_{1} \over 2}}q^{-{\epsilon_{n} \over 2}}\ e_{n}, ~~~\hat t^{0-}_{n1} = - 2 \sinh i\mu\  q^{{1\over 2}} q^{-{\epsilon_{1} \over 2}}q^{-{\epsilon_{n} \over 2}}\ f_{n} \non\\ && t_{ii} =\hat t_{ii}=(t_{ii}^{-})^{-1} =(\hat t_{ii}^{-})^{-1}=q^{\epsilon_{i}}. \label{q2} \ee The interesting feature of the elements $t_{ij}$, $\hat t_{ij}$ is that they form, as shown explicitly below, very simple coproduct expressions compatible with (\ref{cop}) $\Delta: {\cal A} \to {\cal A} \otimes {\cal A}$ i.e., \be \Delta(t_{ij}) = \sum_{k=i}^{j}t_{kj}\otimes t_{ik}, ~~~\Delta(\hat t_{ji}) = \sum_{k=i}^{j} \hat t_{jk}\otimes  \hat t_{ki},  ~~~\Delta(\hat t^{-}_{ji}) = \sum_{k=i}^{j}  t^{-}_{ki}\otimes t^{-}_{jk}  ~~~i<j \label{ns2}  \ee (the coproduct $\Delta(\hat t_{ij}^{-})$ is omitted for brevity and we never actually use it in the present analysis). The coproducts (\ref{ns2}) are restricted to the non-affine case, whereas for the elements $t_{1n}^0,\ t_{n1}^{0}$, which are related to the affine generators the coproducts read as: \be  && \Delta(y) = t_{11} \otimes y +y \otimes t_{nn}, ~~~y \in \{t_{1n}^{0},\ t_{n1}^{0} \} \label{ns3} \ee (similarly we omit $\Delta(t_{n1}^{0-})$). 
    
We shall now derive  the coproducts of the elements ${\cal E}_{ij}$ (\ref{q1})
and subsequently $t_{ij}$ (\ref{q2}).\\
The following summation identities, which may be shown by induction, will be useful for both appendices A and C:\\
\\
$~~\sum_{j=i}^{m}\ \sum_{k=i}^{j} f_{jk} = \sum_{j=k}^{m}\ \sum_{k=i}^{m}f_{jk}~~$,  $~~\sum_{j=i}^{m}\ \sum_{k=j}^{m} f_{jk} = \sum_{j=i}^{k}\ \sum_{k=i}^{m}f_{jk}.~~~~$  (i)\\
\\
Define also: $~~w = 2\ \sinh i \mu $.

We shall first show that: \be \Delta({\cal E}_{ij}) = q^{-{\epsilon_{j} -\epsilon_{i} \over 2}} \otimes {\cal E}_{ij} + {\cal E}_{ij} \otimes q^{{\epsilon_{j} -\epsilon_{i} \over 2}} + q^{-{1\over 2}} w\ \sum_{k =j+1}^{i-1} q^{-{\epsilon_{j} -\epsilon_{k} \over 2}} {\cal E}_{ik} \otimes q^{{\epsilon_{k} -\epsilon_{i} \over 2}} {\cal E}_{kj}, ~~~i-j > 1. \label{copa} \ee We shall proceed with the proof by induction.\\
{\bf 1.} The first step is to show (\ref{copa}) for $i-j =2$, indeed from the definition (\ref{q1}) we have that (we use ${\cal E}_{ij}$, $~~i>j$ for the proof, but the process is similar for $\hat {\cal E}_{ij}$ and ${\cal E}_{ij}$, $i<j$)
\be \Delta({\cal E}_{i+2\ i}) = \Delta({\cal E}_{i+2\ i+1})\ \Delta({\cal E}_{i+1\ i}) -q^{-1}\Delta({\cal E}_{i+1\ i})\ \Delta({\cal E}_{i+2\ i+1}) \label{ap1} \ee but the coproduts of ${\cal E}_{i+i\ i}=f_{i}$ are given by (\ref{cop}) so by substituting expressions from (\ref{cop}) in (\ref{ap1}) and also taking into account relations (\ref{1}) and (\ref{chev}) we obtain after some algebra \be \Delta({\cal E}_{i+2\ i}) = q^{-{\epsilon_{i} -\epsilon_{i+2}\over 2}} \otimes {\cal E}_{i+2\ i} +{\cal E}_{i+2\ i} \otimes q^{{\epsilon_{i} -\epsilon_{i+2} \over 2}} +q^{-{1 \over 2}} w\ q^{-{\epsilon_{i} -\epsilon_{i+1} \over 2}} {\cal E}_{i+2\ i+1} \otimes q^{{\epsilon_{i+1} -\epsilon_{i+2} \over 2}}{\cal E}_{i+1\ i}. \ee
{\bf 2.} In the second step we assume that (\ref{copa}) is true for all $i-j <m$, $~m>2$ and\\
{\bf 3.} we shall prove that (\ref{copa}) is valid for $i-j =m$. The proof is straightforward, from (\ref{q1}) by substituting $\Delta({\cal E}_{kj})$ and $\Delta({\cal E}_{ik})$ from (\ref{copa}) we have that $\Delta({\cal E}_{ij})$, $~i-j=m$ is : \be && {1\over m-1} \sum_{k=j+1}^{i-1}\Big [ \Big (q^{-{\epsilon_{k} -\epsilon_{i} \over 2}} \otimes {\cal E}_{ik} + {\cal E}_{ik} \otimes q^{{\epsilon_{k} -\epsilon_{i} \over 2}} + q^{-{1\over 2}} w \sum_{l =k+1}^{i-1} q^{-{\epsilon_{k} -\epsilon_{l} \over 2}} {\cal E}_{il} \otimes q^{{\epsilon_{l} -\epsilon_{i} \over 2}} {\cal E}_{lk}\Big ) \non\\ && \times \Big (q^{-{\epsilon_{j} -\epsilon_{k} \over 2}} \otimes {\cal E}_{kj} + {\cal E}_{kj} \otimes q^{{\epsilon_{j} -\epsilon_{k} \over 2}} + q^{-{1\over 2}} w \sum_{l' =j+1}^{k-1} q^{-{\epsilon_{j} -\epsilon_{l'} \over 2}} {\cal E}_{kl'} \otimes q^{{\epsilon_{l'} -\epsilon_{k} \over 2}} {\cal E}_{l'j} \Big)  \non\\ && -q^{-1}\Big( \ldots \Big) \Big ]= \ldots \ldots \ee then by using (\ref{1}), (\ref{chev}) and (i) we conclude 
\be && q^{-{\epsilon_{j} -\epsilon_{i} \over 2}} \otimes {\cal E}_{ij} + {\cal E}_{ij} \otimes q^{{\epsilon_{j} -\epsilon_{i} \over 2}} + q^{-{1\over 2}} {w \over m-1} \sum_{k =j+1}^{i-1} q^{-{\epsilon_{j} -\epsilon_{k} \over 2}} {\cal E}_{ik} \otimes q^{{\epsilon_{k} -\epsilon_{i} \over 2}} {\cal E}_{kj} \non\\ &+&  q^{-{1\over 2}} {w \over m-1} \sum_{l' =j+1}^{i-1} \sum_{k =l'+1}^{i-1}q^{-{\epsilon_{j} -\epsilon_{l'} \over 2}}\Big  ({\cal E}_{ik}{\cal E}_{kl'} -q^{-1}{\cal E}_{kl'}{\cal E}_{ik}\Big )\otimes q^{{\epsilon_{l'} -\epsilon_{i} \over 2}} {\cal E}_{l'j} \non\\ &+&  q^{-{1\over 2}} {w \over m-1} \sum_{l =j+1}^{i-1} \sum_{k=j+1}^{l-1} q^{-{\epsilon_{j} -\epsilon_{l} \over 2}} {\cal E}_{il} \otimes q^{{\epsilon_{l} -\epsilon_{i} \over 2}} \Big ({\cal E}_{lk}{\cal E}_{kj} -q^{-1}{\cal E}_{kj}{\cal E}_{lk}\Big ). \label{up}
\ee Finally by multiplying and dividing the last two terms in the latter expression by $( i-l'-1)$ and $(l-j-1)$ respectively and by virtue of (\ref{q1}), after adding the last three terms in (\ref{up}),  we end up to (\ref{copa}) for $i-j =m$.\finproof \\ The proof is analogous for $\Delta(\hat {\cal E}_{ij})$, i.e.  \be \Delta(\hat {\cal E}_{ij}) = q^{-{\epsilon_{i} -\epsilon_{j} \over 2}} \otimes \hat {\cal E}_{ij} + \hat {\cal E}_{ij} \otimes q^{{\epsilon_{i} -\epsilon_{j} \over 2}}+ q^{-{1\over 2}} w\ \sum_{k =i+1}^{j-1} q^{-{\epsilon_{i} -\epsilon_{k} \over 2}} \hat {\cal E}_{kj} \otimes q^{{\epsilon_{k} -\epsilon_{j} \over 2}} \hat {\cal E}_{ik}, ~~~j-i > 1. \ee 
Now we can show the coproduct expressions for $t_{ij}$ and $\hat t_{ij}$ (\ref{ns2}). From equation (\ref{q2}) it follows for $~i<j~$ that \be \Delta(t_{ij}) = q^{-{1\over 2}}w\ \Big ( q^{{ \epsilon_{i} +  \epsilon_{j} \over 2}} \otimes q^{{ \epsilon_{i} +  \epsilon_{j} \over 2}} \Big ) \Big (q^{-{ \epsilon_{i} -  \epsilon_{j}\over 2}} \otimes {\cal E}_{ji} + {\cal E}_{ji} \otimes q^{{ \epsilon_{i} -  \epsilon_{j} \over 2}} + q^{-{1\over 2}} w\ \sum_{k=i+1}^{j-1}q^{-{\epsilon_{i} -\epsilon_{k} \over 2}}{\cal E}_{jk} \otimes q^{{\epsilon_{k} -\epsilon_{j} \over 2}}{\cal E}_{ki}\Big )\ee and by virtue of (\ref{q2}) we conclude 
\be \Delta(t_{ij}) = \sum_{k=i}^{j} t_{kj} \otimes t_{ik}, ~~~i<j. \label{copab}  \ee A similar proof holds for $\Delta(\hat t_{ij})$.

\section{Appendix}

We shall compute in this appendix the quantity $tr_{0}\ M_{0}\ U_{N0}$. Recall that 
$U_{N0} =U_{0N}(q\to q^{-1})$ (\ref{sol}) then
\be M_{0}\ U_{N0} = (\sum_{i=1}^{n} q^{(n-2i+1)}\hat e_{ii} \otimes {\mathbb I} )\sum_{k\neq l =1}^{N} (\hat e_{kl} \otimes \hat e_{lk} -q^{sgn(k-l)}\hat e_{kk}\otimes \hat e_{ll} ). \ee Recalling also the basic property: $~~ \hat e_{ij}\ \hat e_{kl} = \delta_{kj}\ \hat e_{il}~~$  we obtain
\be  M_{0}\ U_{N0} =  \sum_{k\neq l =1}^{n} q^{(n-2k +1)}( \hat e_{kl} \otimes \hat e_{lk} -q^{sgn(k-l)}\hat e_{kk}\otimes \hat e_{ll}). \label{mu} \ee
Now we need to compute the trace of (\ref{mu}) which reads:
\be tr_{0}\ M_{0}\ U_{N0} &=& - \sum_{k\neq l}q^{(n-2k+1)}q^{sgn(k-l)} \hat e_{ll} \non\\ &=& - \sum_{k=1}^{l-1}\sum_{l=1}^{n}q^{n-2k} \hat e_{ll} - \sum_{k=l+1}^{n}\sum_{l=1}^{n}q^{n-2k+2} \hat e_{ll} = \ldots \non\\ &=& -{\sinh  i \mu(n-1) \over \sinh i \mu}  \sum_{l=1}^{n} \hat e_{ll} \label{mu1} \ee but $\sum_{l=1}^{n} \hat e_{ll} = \rho(1)$.\\
Finally the constant $c_{0}$ appearing in (\ref{c}) can be derived, indeed from $M$ in the homogeneous gradation (\ref{M}) it follows $~~tr_{0}\ M_{0}= {\sinh i \mu n \over \sinh i \mu}~~$ and finally via (\ref{c}), (\ref{mu1}) we conclude that 
\be c_{0}=-{\sinh i \mu (n-1) \over \sinh i \mu n}. \ee 

\section{Appendix} 

In this appendix we present explicit expressions for the
$L$ coproducts for the charges (\ref{genb}), (\ref{genb2}). In particular, taking into account the expressions (\ref{genb}), (\ref{genb2}) ($t_{ij}, \hat t_{ij} \to
\Delta^{(L)}(t_{ij}),\Delta^{(L)}(\hat t_{ij})$), the coproducts (\ref{cop}), and the identities (i) defined in Appendix A we can write
\be   \Delta^{(L)}({\cal Q}_{1i}) &=& \sum_{k=i}^{n-1}{\cal Q}_{1k} \otimes \Delta^{(L-1)}(t_{11})\Delta^{(L-1)}(\hat t_{ki}) +e^{i\mu m}\sum_{j=i}^{n-1} \sum_{k=2}^{j} \sum_{l=i}^{j} t_{kj} \hat t_{jl} \otimes \Delta^{(L-1)}(t_{1k}) \Delta^{(L-1)}(\hat t_{li}) \non\\ &-& i t_{11}  t_{nn} \otimes \Delta^{(L-1)}(t_{11}) \Delta^{(L-1)}(\hat t_{ni}) \non\\ \Delta^{(L)}({\cal Q}_{i1}) &=& \sum_{k=i}^{n-1}{\cal Q}_{k1} \otimes \Delta^{(L-1)}(t_{ik})\Delta^{(L-1)}( t_{11}) +e^{i\mu m}\sum_{j=i}^{n-1} \sum_{k=i}^{j} \sum_{l=2}^{j} t_{kj} \hat t_{jl} \otimes \Delta^{(L-1)}(t_{ik}) \Delta^{(L-1)}(\hat t_{l1}) \non\\ &-& it_{nn}  t_{11} \otimes \Delta^{(L-1)}(t_{in} )\Delta^{(L-1)}(t_{11})  \non\\  \Delta^{(L)}({\cal Q}_{11}) &=& \sum_{k=1}^{n-1}{\cal Q}_{1k} \otimes \Delta^{(L-1)}(t_{11})\Delta^{(L-1)}(\hat t_{k1}) +\sum_{k=2}^{n-1}{\cal Q}_{k1}\otimes \Delta^{(L-1)}(t_{1k}) \Delta^{(L-1)}( t_{11}) \non\\ &+&e^{i\mu m}\sum_{j=2}^{n-1} \sum_{k=2}^{j} \sum_{l=2}^{j} t_{kj} \hat t_{jl} \otimes \Delta^{(L-1)}(t_{1k}
 ) \Delta^{(L-1)}(\hat t_{l1}) \non\\ &-&it_{11}  t_{nn} \otimes (\Delta^{(L-1)}(t_{1n}) \Delta^{(L-1)}(t_{11})+ \Delta^{(L-1)}( t_{11}) \Delta^{(L-1)}(\hat t_{n1}) ) \non\\  \Delta^{(L)}({\cal Q}_{nn}) &=& ({\cal Q}_{nn}+2 \cosh 2 i \mu \zeta\ t_{11} t_{nn}) \otimes (\Delta^{(L-1)}  (t_{nn}))^{2} +t_{11} t_{nn}\otimes \Delta^{(L-1)}({\cal Q}_{nn}).  \label{bc2} \ee  Similar expressions may be derived for the $\Delta^{'(L)}$ coproducts via (\ref{perm}), (\ref{cop2}).

\section{Appendix}

It is instructive to present the elements  $(\pi_{\lambda} \otimes \mbox{id}^{\otimes N})\Delta^{'(N+1)}(e_{i})$,  $(\pi_{\lambda} \otimes \mbox{id}^{\otimes N})\Delta^{'(N+1)}(f_{i})$ $i\in\{1, \ldots n-1 \}$ in a matrix form 
\begin{equation} (\pi_{\lambda} \otimes \mbox{id}^{\otimes N})\Delta^{'(N+1)}(e_{i})=\left(\begin{array}{cccccc}
E_{i\ i+1}^{(N)} &0  &\cdots &\cdots &\cdots &0\\
\vdots   &\ddots &\ddots &\ddots  &\ddots  &\vdots\\
0 &\ldots &q^{{1\over 2}} E_{i\ i+1}^{(N)} &\underbrace{(H_{i}^{(N)})^{-{1 \over 2}}}_{\mbox{$(i,\ i+1)$}} &\ldots  &0\\
0 &\ldots &0   &q^{-{1\over 2}} E_{i\ i+1}^{(N)} &\ldots  &0\\
\vdots &  &\ddots & \ddots & \ddots     & \vdots\\
0  & \cdots &\cdots &\cdots    & 0 & E_{i\ i+1}^{(N)}
\end{array}\right) \label{form1}
\end{equation}
\begin{equation} (\pi_{\lambda} \otimes \mbox{id}^{\otimes N})\Delta^{'(N+1)}(f_{i})=\left(\begin{array}{cccccc}
E_{i+1\ i}^{(N)} &0  &\cdots &\cdots &\cdots &0\\
\vdots & \ddots  & \ddots  & \ddots  & \ddots  & \vdots\\
0 &\ldots   &q^{{1\over 2}} E_{i+1\ i}^{(N)} &0 &\ldots &0\\
0 &\ldots   &\underbrace{(H_{i}^{(N)})^{-{1 \over 2}}}_{\mbox{$(i+1,\ i)$}} &q^{-{1\over 2}} E_{i+1\ i}^{(N)} &\ldots  &0\\
\vdots   &\ddots &\ddots  &\ddots  &\ddots  &\vdots\\
0  &\cdots  &\cdots  &\cdots  &0 & E_{i+1\ i}^{(N)}
\end{array}\right)\label{form2}
\end{equation} and finally \be \Big ((\pi_{\lambda} \otimes \mbox{id}^{\otimes N})\Delta^{'(N+1)}(q^{\epsilon_{i}}) \Big )_{kl} = E_{ii}^{(N)}\ (\delta_{kl} -\delta_{ik}\ \delta_{il}) + q\ E_{ii}^{(N)}\ \delta_{ik}\ \delta_{il}. \label{form2b} \ee Note that the corresponding expressions for the affine generators $e_{n}$ and $f_{n}$ have a similar form with (\ref{form1}), (\ref{form2}) but with the element $e^{\mp 2 \lambda}\ (H_{n}^{(N)})^{-{1 \over 2}}$ being at the positions $(n,\ 1)$ and $(1,\ n)$, and the factors $q^{\pm {1\over 2}}$ multiplying the first and the last of the diagonal elements $E^{0(N)}_{n1}=\pi_{0}^{\otimes N} (\Delta^{(N)}(e_{n}))$ and $E^{0(N)}_{1n}=\pi_{0}^{\otimes N}(\Delta^{(N)}(f_{n}))$ respectively.
   
We can also  easily write down the expression for $(\pi_{\lambda} \otimes \mbox{id}^{\otimes N})\Delta^{'(N+1)}({\cal Q}_{nn})$ for any $n$, i.e. 
\begin{equation} (\pi_{\lambda} \otimes \mbox{id}^{\otimes N})\Delta^{'(N+1)}({\cal Q}_{nn})=\left(\begin{array}{cccccc}
{\cal T}^{+(N)}_{nn} &0 &\cdots &\cdots &\cdots &-i e^{2\lambda}q w E_{11}^{(N)} E_{nn}^{(N)} \\
\vdots & \ddots  & \ddots  & \ddots  & \ddots  & \vdots\\
0 &\ldots   &{\cal T}^{+(N)}_{nn} &0 &\ldots &0\\
0 &\ldots   &0 &{\cal T}^{+(N)}_{nn} &\ldots  &0\\
\vdots   &\ddots &\ddots  &\ddots  &\ddots  &\vdots\\
-i e^{-2  \lambda}q w E_{11}^{(N)} E_{nn}^{(N)} &\cdots &\cdots &\cdots &0 & q^{2}{\cal T}^{+(N)}_{nn}
\end{array}\right).\label{form4}
 \end{equation}  
Let us finally give explicit expressions of representations of the boundary charges associated to the $U_{q}(\widehat{gl_{3}})$ case, i.e. from (\ref{eval}), (\ref{bc2}) we get
\be &&(\pi_{\lambda} \otimes \mbox{id}^{\otimes N})\Delta^{'(N+1)}({\cal Q}_{11})= \left(\begin{array}{ccc}
q^2 {\cal T}^{+(N)}_{11} &w q {\cal T}^{+(N)}_{12} &-i w q E_{11}^{(N)} E_{33}^{(N)} \\
w q {\cal T}^{+(N)}_{21} &{\cal T}^{+(N)}_{11}+ e^{i\mu m} w^{2}(E_{22}^{(N)})^{2} &0 \\
-i w qE_{11}^{(N)} E_{33}^{(N)}  &0 &{\cal T}^{+(N)}_{11}\\
\end{array}\right) \non\\ &&(\pi_{\lambda} \otimes \mbox{id}^{\otimes N})\Delta^{'(N+1)}({\cal Q}_{12})= \left(\begin{array}{ccc}
q {\cal T}^{+(N)}_{12} &0 &0 \\
e^{i\mu m} w (E^{(N)}_{22})^2 &q {\cal T}^{+(N)}_{12} &-i wE_{11}^{(N)} E_{33}^{(N)}  \\
0  &0 &{\cal T}^{+(N)}_{12}\\
\end{array}\right)   \non\\  &&(\pi_{\lambda} \otimes \mbox{id}^{\otimes N})\Delta^{'(N+1)}({\cal Q}_{21})=\left(\begin{array}{ccc}
q {\cal T}^{+(N)}_{21} &e^{i\mu m} w (E^{(N)}_{22})^2 &0 \\
0 &q {\cal T}^{+(N)}_{21} &0  \\
0  &-i wE_{11}^{(N)} E_{33}^{E(N)} &{\cal T}^{+(N)}_{21}\\
\end{array}\right). \label{form3}  \ee

\end{document}